\documentclass[journal]{IEEEtran}

\usepackage[pdftex]{graphicx}
\graphicspath{Figures/}

\usepackage{amsmath}
\usepackage{amssymb}

\usepackage{tikz}
\usepackage{verbatim}
\usetikzlibrary{arrows,shapes,fit,scopes,calc,matrix,positioning,decorations.pathmorphing,shapes.geometric}

\tikzset{
    vertex/.style={rectangle,draw,minimum width=6em},
    edge/.style={->,> = latex'},
    cvertex/.style={circle,draw,minimum width=0.3em,inner sep=1},
    lcvertex/.style={circle,draw,minimum width=2em,inner sep=1}    
}

\newcommand{\Figure}{Fig.~}

\newtheorem{defn}{Definition}
\newtheorem{example}{Example}
\begin{document}
%
\title{Residual Generation Using Physically-Based Grey-Box Recurrent Neural Networks For Engine Fault Diagnosis}
%
%
%

\author{Daniel Jung
\thanks{D. Jung is with the Department
of Electrical Engineering, Link\"{o}ping University, Link\"{o}ping,
Sweden e-mail: daniel.jung@liu.se.}
}
%
%

\markboth{Preprint}%
{}


%



\maketitle

\begin{abstract}
Data-driven fault diagnosis is complicated by unknown fault classes and limited 
training data from different fault realizations. In these situations, conventional 
multi-class classification approaches are not suitable for fault diagnosis. One solution is 
the use of anomaly classifiers that are trained using only nominal data. Anomaly classifiers 
can be used to detect when a fault occurs but give little information about its root cause. 
Hybrid fault diagnosis methods combining physically-based models and available training data 
have shown promising results to improve fault classification performance and identify unknown 
fault classes. Residual generation using grey-box recurrent neural networks can be used for 
anomaly classification where physical insights about the monitored system are incorporated into 
the design of the machine learning algorithm. In this work, an automated residual design is 
developed using a bipartite graph representation of the system model to design grey-box 
recurrent neural networks and evaluated using a real industrial case study. Data from 
an internal combustion engine test bench is used to illustrate the potentials of 
combining machine learning and model-based fault diagnosis techniques.  
\end{abstract}


%
\IEEEpeerreviewmaketitle

\section{Introduction}

Fault diagnosis of industrial systems is about monitoring the system 
health including detection of occurring faults and identifying their root 
cause. One of the main principles of fault diagnosis is to detect inconsistencies 
between sensor data from the monitored system and predictions computed 
based on a model of the system behavior. Two common approaches to compute 
predictions are based on physically-based or data-driven models. 

Model-based diagnosis relies on physically-based models that are derived 
based on first principle physics. Model parameters can be derived from 
physical properties of the system to be monitored, referred to as white-box modeling, 
or estimated from data, referred to as grey-box modeling \cite{ljung1999system}.

Data-driven models rely on representative training data to select and train a 
general-purpose model structure that best captures the multi-variate information 
in training data. The parameters and model structure of data-driven models do 
seldom have a physical interpretation with respect to the modeled system and 
are therefore referred to as black-box models \cite{ljung1999system}. 

Developing sufficiently accurate physically-based models for residual generation 
is a time-consuming process and requires expert knowledge about the system to 
be modeled, especially for complex or large-scale systems. This have motivated the 
use of data-driven modeling and machine learning to design diagnosis systems 
\cite{qin2012survey}. However, collecting representative training data for fault diagnosis 
applications is a difficult task. Many different types of faults could occur in the 
system and each type of fault can have many different realizations due to 
varying operating conditions and fault magnitudes \cite{jung2018combining}. 
Certain fault types can be difficult to collect data from and there could be faults 
that are not considered when designing the diagnosis system, resulting in 
imbalanced training data and unknown fault classes that are not represented in 
training data, which complicates conventional closed set multi-class data-driven 
fault classification \cite{dong2017method,sankavaram2015incremental}. 

Instead of using data-driven models with a general-purpose model structure, there 
are benefits when selecting a model structure that captures the physical 
relationship between input and output variables 
\cite{acuna1999comparison}. It is possible to reduce the model parameter space 
and the number of input variables by identifying which variables have sufficient 
information for regression, i.e. to capture the behavior of a predicted variable. 
If the model structure is given, less training data are needed to model the 
remaining non-linearities which also reduces the risk of overfit. A physically-based 
model structure also improves model interpretability since different parts of the model 
correspond to different system components \cite{pintelas2020grey}. 
Another potential benefit is to achieve similar fault isolation capabilities as model-based 
residuals \cite{jung2018combining}. By including physical insights in 
data-driven anomaly classifiers it is possible to identify the root cause of 
unknown fault classes, for example, by mapping triggering residuals to components that are 
modeled in each residual generator \cite{jung2019isolation}.  

Recurrent neural networks (RNN) are powerful black-box models able to capture the behavior 
of non-linear dynamic systems \cite{arsie2006procedure}. Neural networks have a flexible 
model structure making them applicable in many different applications. However, a general 
drawback of general-purpose models, such as neural networks, is that they can contain 
lots of parameters to fit in order to model the behavior of the system which requires a significant 
amount of training data to avoid overfitting \cite{aggarwal2018neural}. 
Utilizing physical insights about the system to be monitored can help select a neural network 
model structure that resembles the physical system. Designing grey-box 
RNN based on physical models for residual generation has been proposed in, e.g., 
\cite{pulido2019state}. In \cite{jung2019isolation}, a simulation study indicates that 
grey-box neural networks can identify the root cause of unknown fault classes. However, 
to make this useful for fault diagnosis in industrial systems, it is relevant to investigate 
how to systematically design and train grey-box RNN to achieve certain fault classification 
properties.  
\section{Problem Statement}

Designing grey-box RNN for residual generation is a promising approach 
to combine physical insights about the system to be monitored and machine learning 
for fault diagnosis. Since faults are rare events, it is likely that training data are imbalanced 
and consists mainly of data from fault-free system operation. Residual generators are 
anomaly classifiers meaning that training only requires fault-free data to detect abnormal 
system behavior caused by faults. By designing different residual generators that are 
sensitive to different faults, it is possible to isolate faults by analyzing the resulting fault 
patterns. 

In this work, the objective is to develop a diagnosis system combining model-based and 
machine learning methods applied to an internal combustion engine test bench. A set of 
data-driven residual generators are designed using grey-box RNN and trained using only 
nominal training data. An automated design process is developed where the RNN models 
are generated based on a structural model representing the physical insights about the 
system model structure. 

Structural models are bi-partite graphs that represent the qualitative relationship 
between different model variables and can be used even though the analytical 
relationship is not completely known \cite{cassar1997structural}. 
The usefulness of structural models for fault diagnosis analysis and diagnosis system 
design have been shown in, e.g., \cite{pulido2004possible, krysander2007efficient}, 
and can be used for, for example, fault isolability analysis, residual generation, and 
sensor placement \cite{frisk2017toolbox}. 

To evaluate the proposed method, an internal combustion engine test bench is used 
as a case study. Both training and validation data are collected from the test bench when the 
engine is working in transient operation. A structural model of the air path through 
the engine is used to represent the available model information that is used to 
design different grey-box RNN modeling different parts of the engine. 
\section{Related Research}

Data-driven classification that considers both known and unknown classes is referred to 
as open set classification \cite{scheirer2012toward}. Different approaches to open set 
classification have been proposed, see e.g. \cite{jung2018combining} and \cite{atoui2019single}. 
Even though these mentioned work are able to identify when data belong to unknown
fault classes, it is not trivial how to identify the true class label. In \cite{jung2018combining},
model-based residuals are used to compute fault hypotheses based on the fault 
sensitivity of each residual generator and a set of anomaly classifiers are used to 
rank the known fault classes.    

Neural networks have been used for fault diagnosis in many different applications 
\cite{bernieri1994neural}. Feature extraction using neural networks is proposed in, 
e.g. \cite{zhang2020knowledge}. In \cite{bidarvatan2014grey}, neural networks are used to 
develop a grey-box simulation model of an HCCI engine. In \cite{witczak2006advances}, 
neural networks and genetic programming are used for non-linear residual design for fault 
diagnosis of an industrial valve actuator. A hybrid system identification approach combining 
model-based and neural networks is proposed in \cite{lu2019model} where a weighted 
prediction is computed from the model-based and data-driven models. In \cite{rahimilarki2018grey}, 
neural networks are used for fault diagnosis of a wind turbine.  
 
The benefits of bridging and combining physically-based fault diagnosis methods and 
machine learning have been discussed in, e.g., \cite{tidriri2016bridging}. The connections 
between different neural network structures and ordinary differential equations have been 
analyzed in, e.g. \cite{lu2018beyond} and \cite{chen2018neural}. Including physical insights 
about the system to design grey-box neural networks have been proposed in previous 
works, such as \cite{pulido2019state,wu2016membership} and \cite{hofmann2019comparison}. 
In \cite{choudhary2020physics}, hamiltonian dynamics are incorporated in the neural 
network structure. Grey-box RNN, based on state-space neural networks \cite{zamarreno2000state}, 
are proposed in \cite{pulido2019state} for residual generation to perform fault diagnosis 
of an evaporator in a sugar beet factory. With respect to previous work, an automated design 
process of grey-box RNN for residual generation is developed here combining physically-based 
structural models and deep learning techniques applied to an automotive case study. 
\section{Artificial Neural Networks}

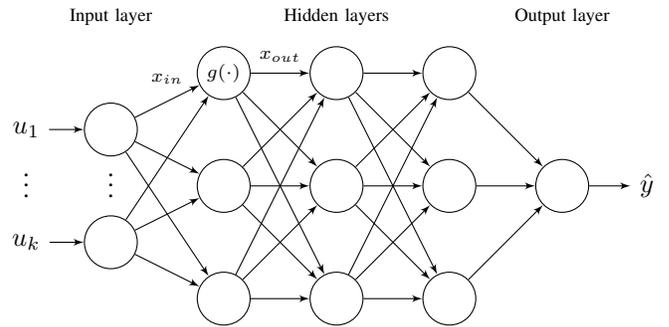
\begin{figure}[!t]
\centering
\begin{tikzpicture}[scale=0.75]
\node[] (1u1) at (0.5,0) {$u_1$};
\node[] (1u2) at (0.5,-2) {$u_k$};
\node[] (1dots) at (0.5,-0.85) {$\vdots$};
\node[] (2dots) at (2,-0.85) {$\vdots$};

\node[lcvertex] (1l1) at (2,0) {};
\node[lcvertex] (1l2) at (2,-2) {};

\draw (2,2) node[]{\scriptsize Input layer};

\node[lcvertex] (2l1) at (4,1) {\scriptsize $g(\cdot)$};
\node[lcvertex] (2l2) at (4,-1) {};
\node[lcvertex] (2l3) at (4,-3) {};

\draw (3.0,0.9) node[]{\scriptsize $x_{in}$};
\draw (5.0,1.3) node[]{\scriptsize $x_{out}$};

\node[lcvertex] (3l1) at (6,1) {};
\node[lcvertex] (3l2) at (6,-1) {};
\node[lcvertex] (3l3) at (6,-3) {};

\draw (6,2) node[]{\scriptsize Hidden layers};

\node[lcvertex] (4l1) at (8,1) {};
\node[lcvertex] (4l2) at (8,-1) {};
\node[lcvertex] (4l3) at (8,-3) {};

\node[lcvertex] (5l1) at (10,-1) {};
\node[] (6y1) at (11.5,-1) {$\hat{y}$};

\draw (10,2) node[]{\scriptsize Output layer};

\draw[edge] (1u1) -- (1l1);
\draw[edge] (1u2) -- (1l2);

\foreach \x in {1l1, 1l2} {%
    \foreach \y in {2l1, 2l2, 2l3} {%
        \draw[edge] (\x) -- (\y);
    }
}  

\foreach \x in {2l1, 2l2, 2l3} {%
    \foreach \y in {3l1, 3l2, 3l3} {%
        \draw[edge] (\x) -- (\y);
    }
}  

\foreach \x in {3l1, 3l2, 3l3} {%
    \foreach \y in {4l1, 4l2, 4l3} {%
        \draw[edge] (\x) -- (\y);
    }
} 

\foreach \x in {4l1, 4l2, 4l3} {%
    \draw[edge] (\x) -- (5l1);
} 

\draw[edge] (5l1) -- (6y1);
\end{tikzpicture}
\caption{An example of an artificial neural network with $k$ inputs and one output.}
\label{fig:ann}
\end{figure}

An artificial neural network models the relationship between a set of 
inputs $u \in \mathbb{R}^{n_u}$ and outputs $y \in \mathbb{R}^{n_y}$ 
using a computational graph, as illustrated in \Figure\ref{fig:ann}, where 
each node represents a non-linear operation on the inputs to the node, $x_{in}$, 
usually in the form: 
\begin{equation}
x_{out} = g(a^T x_{in} + b)    
\end{equation}
where $g(\cdot)$ is a non-linear function, called 
an activation function, $a$ is vector of weights, 
and $b$ is a bias term. Some common activation functions are the 
rectified linear unit (ReLU) 
\begin{equation}
g(\xi) = \max\left(0,\xi\right) \nonumber
\end{equation}
and different sigmoid functions, such as 
the logistic function or arc tangent function \cite{aggarwal2018neural}. 

Neural networks are commonly designed such that the nodes are 
organized in different layers where the inputs to each node are the 
output of nodes in the previous layer and the nodes in the same 
layers have the same activation function. The first layer is referred 
to as the input layer, where data is fed into the neural network. The 
last layer is the output layer, which returns the output from the neural 
network model, and the in-between layers are referred to as hidden 
layers. The number of layers denotes the depth and the maximum 
number of nodes in any layer the width of the neural network. 

Training of neural networks is performed by defining a cost function, 
e.g. mean square error $\sum(y - \hat{y})^2$ for regression problems, 
and updating the model parameters by computing gradients using 
back-propagation and automatic differentiation \cite{aggarwal2018neural}. 
Training neural networks is a non-linear optimization problem that 
can be computationally demanding since these models can have a 
large amount of parameters to fit. Different optimization solvers, 
regularization techniques, and training strategies have been proposed to 
improve learning rate, avoid overfit and reduce the risk of getting stuck in 
bad local minima \cite{aggarwal2018neural}.

\subsection{Recurrent Neural Networks}

Recurrent neural networks are used to model dynamic systems and 
time-series data. Internal states are modeled by in the neural network by 
duplicating the network for each time instance and add connections in 
some nodes between different time steps, similar as state variables in a 
state-space model \cite{aggarwal2018neural}. 

\section{Model-based diagnosis and structural modeling}

Model-based diagnosis uses physically-based models of the system 
to be monitored to compute residuals $r$ that compare model 
predictions $\hat{y}$ and sensor data $y$ to detect inconsistencies, 
as illustrated in \Figure\ref{fig:residual}. An advantage of model-based 
diagnosis, with respect to data-driven methods, is that it is possible to identify the 
root cause of unknown faults by designing residual generators where the 
effects of certain faults are decoupled. One efficient approach to analyze 
large-scale non-linear models is called structural methods, see for example 
\cite{krysander2007efficient}. In this work, structural methods will be used 
to find and generate neural network models for residual generation. Here, 
an introduction to model-based diagnosis and structural models is presented.

\tikzstyle{block} = [rectangle, draw, fill=blue!10, 
    text width=4em, text centered, minimum height=3.0em]
\tikzstyle{wide_block} = [rectangle, draw, fill=blue!10, 
    text width=6.0em, text centered, minimum height=2.5em]
\tikzstyle{line} = [draw, -latex']
\tikzstyle{sum} = [draw, fill=blue!10, circle, node distance=1cm]
\begin{figure}[h!]
\centering
  \begin{tikzpicture}[node distance=0.2cm and 0.2cm]
	\node [wide_block] (system) {\includegraphics[width=55pt]{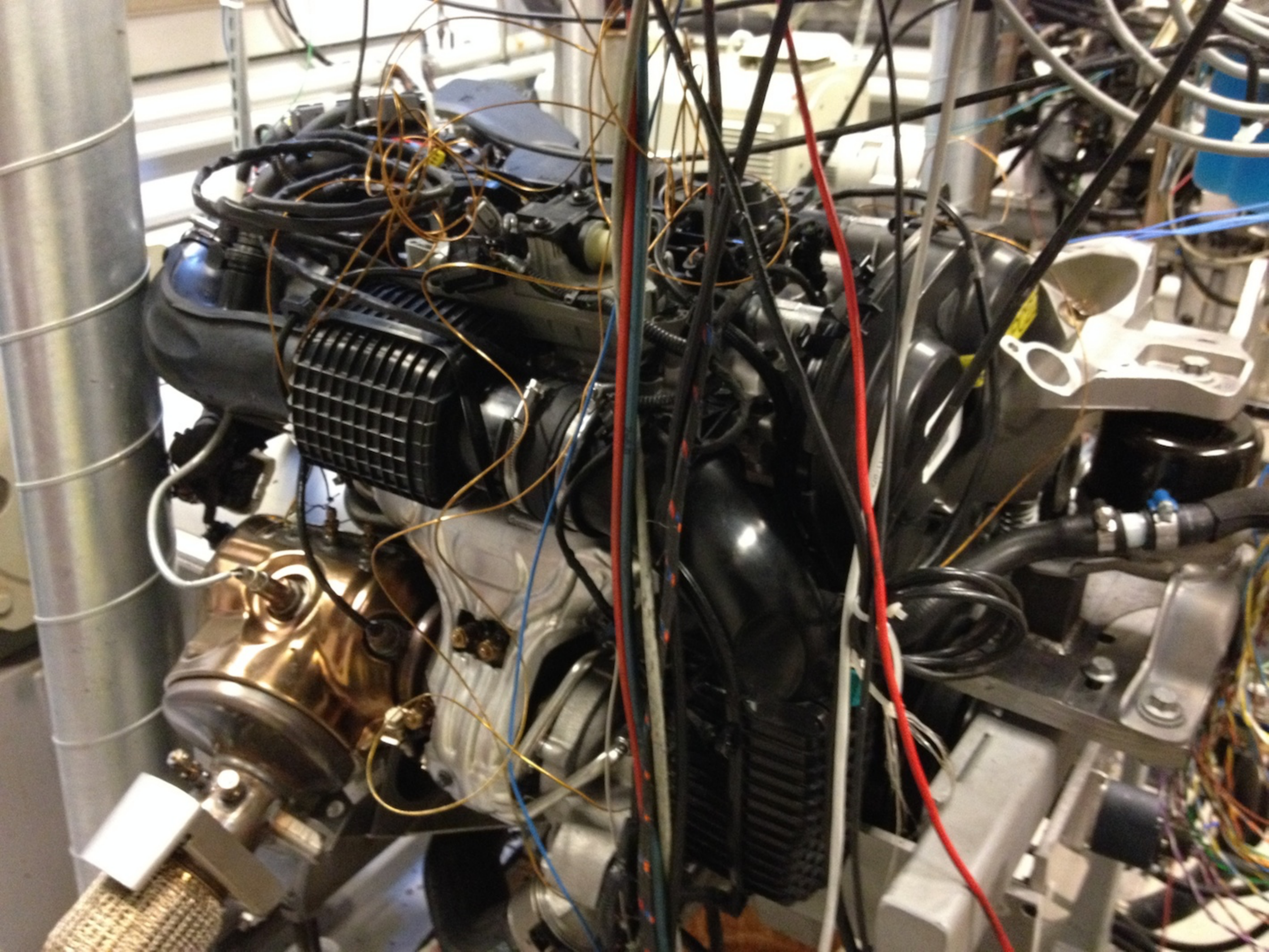}};
	\node [block, below= of system] (model) {\scriptsize Model \\ \vspace{0.1cm} $\begin{aligned} \dot{\hat{x}} &= g(\hat{x}, u) \\ \hat{y} &= h(\hat{x},u) \end{aligned}$};
	\draw ($(system.east) + (1.5,0.0)$) node[sum] (sum) {+};
	\draw[line, ultra thick]  ($(system.west) + (-1.0,0.40)$) node[above]{\scriptsize $f_t$} -- ($(system.west) + (-0.4,0.40)$) --  ($(system.west) + (0.0,0.40)$);
	\draw[line, ultra thick]  ($(system.west) + (-1.0,-0.40)$) node[above]{\scriptsize $u_t$} -- ($(system.west) + (-0.75,-0.40)$) --  ($(system.west) + (0.0,-0.40)$);
	\draw[line, ultra thick]  ($(system.west) + (-0.5,-0.40)$) |-  ($(model.west)$);
	\draw[line, ultra thick] (system)  -- node[above]{\scriptsize $y_t$} (sum);
	\draw[line, ultra thick] (model)  -- ($(model.east) + (0.75,0.0)$) -- ($(model.east) + (1.25,0.0)$) node[above]{\scriptsize $\hat{y}_t$} -| (sum);
	\draw[line, ultra thick] (sum)  -- node[above]{\scriptsize $r_t$} ($(sum.east) + (1.00,0.0)$);
	
	\draw[] ($(sum.east) + (-0.60,-0.60)$) node{$-$};
	  \end{tikzpicture}
  \caption{An example of a residual $r_t$ comparing measurements from the system $y_t$ with model predictions $\hat{y}_t$.}
  \label{fig:residual}
\end{figure}
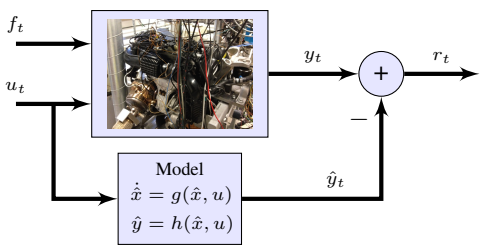

\subsection{Fault detection and isolation}
A residual generator models the nominal 
system behavior where a significant deviation in the residual output 
implies that a fault has occurred. Thus, a residual can be interpreted 
as an anomaly classifier modeling data from the fault-free class \cite{jung2019isolation}. 

By designing residual generators based on different sub-models, it is 
possible to decouple the effects of different faults on the residual outputs 
\cite{jung2018combining}. 
\begin{defn}
A residual $r_k$ is said to be (ideally) sensitive to a fault $f_i$ if 
$f_i \neq 0$ implies that $r_k \neq 0$. 
\end{defn}
If a residual $r_k$ is not sensitive to a fault $f_i$, the fault is said to be 
\emph{decoupled} in that residual. 

Even though the set of residual generators is a set of anomaly classifiers, 
their fault sensitivities can be used to identify the root cause by analyzing 
their fault sensitivities. 
\begin{defn}
A fault $f_i$ is isolable from fault $f_j$ if there is a residual $r_k$ that 
is sensitive to $f_i$ but not $f_j$. 
\end{defn}
The fault sensitivities of a set of residuals can be summarized in a 
fault signature matrix where an X in position $(i,j)$ means that 
residual $r_i$ is sensitive to fault $f_j$. By analyzing the fault sensitivities 
of the set of residuals that are deviating from their nominal behavior, 
a set of fault hypotheses,  also called diagnosis candidates, can be 
computed \cite{de1987diagnosing}. 

\subsection{Change detection}
One of the simplest methods to detect a change in the nominal residual 
output is to compare it with a threshold $r_t - J > 0$ that is tuned to not exceed 
a certain false alarm rate. This approach is not suitable to detect small faults 
while having to fulfill a low false alarm rate. One solution is to use a CUMulative SUm 
(CUSUM) test \cite{page1954continuous}:
\begin{equation}
T_t = \max \left(0, T_{t-1} + r_t - \nu \right), \quad T_0 = 0
\label{eq:cusum}
\end{equation}
where $\nu$ is a tuning parameter. The CUSUM test \eqref{eq:cusum} 
integrates the impact of the fault on the residual output (exceeding the parameter $\nu$) 
over time and a fault is detected by tuning a threshold $J$ such that 
$T_t - J > 0$. This allows for detection of smaller faults without 
increasing the risk of false alarms by allowing a longer time before detection.

\subsection{Structural models}
Structural analysis can be used to analyze fault diagnosis properties 
of complex non-linear dynamic systems and systematic design of 
diagnosis systems \cite{frisk2017toolbox}. A structural model 
$\mathcal{M} = (\mathcal{E}, \mathcal{X})$ is a bi-partite graph 
describing the relationship between model equations $\mathcal{E} = \{e_1, e_2, ...\}$ 
and variables $\mathcal{X} = \{x_1, x_2, ...\}$, i.e. which variables are 
included in each model equation. The structural model can be 
represented using an incidence matrix where an X in row $(i,j)$
mean that variable $x_j$ is included in equation $e_i$. Model 
variables are partitioned into unknown variables, known variables, 
and fault signals that are used to model how different faults are 
affecting the system.

\begin{figure}[!t]
\centering
\begin{tikzpicture}[scale=1]
\draw[->, ultra thick] (0, 0) -- (5, 0);
\draw[->, ultra thick] (0, 0) -- (0, 5);
\draw[-, fill=gray!40] (2.5, 2) -- (4, 2) -- (4,4) -- (1.5,4) -- (1.5,3) -- (2.5,3) -- cycle;
\draw[dashed] (2.5,4) -- (2.5, 3);
\draw[dashed] (2.5,3) -- (4, 3);
\draw[dashed] (0,2) -- (2.5, 2);
\draw[dashed] (1.5,2) -- (1.5, 0);
\draw[ultra thick] (0,0) rectangle (4,4);
\draw[ultra thick,fill=gray!60] (4,0) rectangle (2.5,2);
\draw[ultra thick,fill=gray!60] (2.5,2) rectangle (1.5,3);
\draw[ultra thick,fill=gray!60] (1.5,3) rectangle (0,4);
\draw (3.25,1) node[]{$\mathcal{M}^+$};
\draw (2,2.5) node[]{$\mathcal{M}^0$};
\draw (0.75,3.5) node[]{$\mathcal{M}^-$};
\draw (1.25,1.25) node[]{0};
\draw (3.25,-0.3) node[]{$\mathcal{X}^+$};
\draw (2,-0.3) node[]{$\mathcal{X}^0$};
\draw (0.75,-0.3) node[]{$\mathcal{X}^-$};
\draw (-0.3,1) node[]{$\mathcal{E}^+$};
\draw (-0.3,2.5) node[]{$\mathcal{E}^0$};
\draw (-0.3,3.5) node[]{$\mathcal{E}^-$};
\end{tikzpicture}
\caption{Dulmage-Mendelsohn decomposition of a model $\mathcal{M} = \left( \mathcal{E}, \mathcal{X} \right)$.}
\label{fig:dm}
\end{figure}
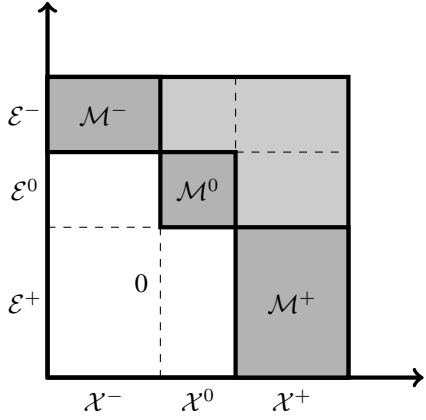

The rows and columns of the incidence matrix can be reorganized 
using the Dulmage-Mendelsohn (DM) decomposition \cite{dulmage1958coverings} to analyze the 
structural redundancy properties of the system \cite{frisk2017toolbox}. 
The DM decomposition partitions a 
structural model into an under-determined part $\mathcal{M}^-$, an 
exactly determined part $\mathcal{M}^0$, and an 
over-determined part $\mathcal{M}^+$ as illustrated in \Figure\ref{fig:dm}. 
The over-determined part of the model has more equations 
than unknown variables and describes the redundant part of 
the model that can be monitored using residual generators 
\cite{krysander2007efficient}. The degree of structural redundancy of a 
model $\mathcal{M}$ is defined as \cite{krysander2007efficient}:
\begin{equation}
\varphi(\mathcal{M}) = |\mathcal{E}^+| - |\mathcal{X}^+|    
\end{equation}
where $| \cdot |$ denotes set cardinality.

A fault that enters the system somewhere modeled in the over-determined 
part, i.e. $e_{f_i} \in \mathcal{M}^+$ where $e_{f_i}$ denotes the model 
equation modeling where the fault $f_i$ manifests in the system, 
is said to be structurally detectable. Similarly, a fault $f_i$ is said to 
be structurally isolable from fault $f_j$ if $e_{f_i} \in \left(\mathcal{M} 
\setminus e_{f_j}\right)^+$, i.e. fault $f_i$ is structurally detectable if $e_{f_i}$
is still in the over-determined part when the equation $e_{f_j}$ is removed 
from the model. In principle, structural fault detectability and 
isolability depend on if it is possible to design a residual generator 
modeling the part of the system where the fault $f_i$ occurs or not.   

Different residual candidates can be found by analyzing different subsets 
of the over-determined part of the model $\mathcal{M}^+$ that are still 
over-determined. By systematically removing equations from the 
over-determined part and analyzing the remaining over-determined part it 
is possible to find all combinations of redundant equation sets, called Partially 
Structurally Over-determined (PSO) sets \cite{krysander2007efficient}. 
\begin{defn}
A structural model $\mathcal{M}$ is called a Partially 
Structurally Over-determined set if $\mathcal{M} = \mathcal{M}^+$.
\end{defn}
A special type of PSO set have degree of redundancy
one, i.e. PSO sets where no subset have redundancy, called Minimally 
Structurally Over-determined sets \cite{krysander2007efficient}. 
\begin{defn}
A PSO set $\mathcal{M} = \mathcal{M}^+$ is called a Minimally 
Structurally Over-determined (MSO) set if $\left(\mathcal{M}\setminus e_k\right)^+ = \emptyset$ for all $e_k \in \mathcal{M}$.
\end{defn}
The MSO sets represent the minimal redundant equation sets that 
can be used for residual generation. MSO sets are interesting from 
a fault isolation perspective since they model a minimal part of the 
system that can be monitored. 

\subsection{Designing Residual Generators Using Computational Graphs}

If one equation is removed from an MSO set, the remaining set is exactly 
determined, meaning that there is an equal amount of unknown variables 
and equations. A matching algorithm can be used to find how to compute all 
unknown variables in the exactly determined set \cite{frisk2012diagnosability}. 
When all unknown variables have been computed in the exactly determined set, 
the redundant equation can be used as a residual equation.

The output from the matching algorithm describing the order of computing the 
unknown variables can be represented as a computational graph. A 
computational graph is a directed graph where nodes either denote a variable 
or a function and edges show how the output of each node are fed as input 
to other nodes, which is illustrated in \Figure\ref{fig:ann}. The order of how the state variables are computed in the computational 
graph will affect its causality. If all the states are computed by integrating 
their derivatives the computational graph is said to have integral causality. 
If the state variable is computed and then differentiated to get its derivative 
the computational graph is said to have derivative causality. If there are 
both states that are integrated and differentiated the computational graph is 
said to have mixed causality \cite{frisk2012diagnosability}. Computational 
graphs of different causalities are illustrated in the following example:
\begin{example}
Consider the following MSO set  
\begin{equation}
\begin{aligned}
e_1: & & \dot{x}_1 &= g_1(u) & & e_4: & & \dot{x}_1 &= \frac{dx_1}{dt}\\
e_2: & & \dot{x}_2 &= g_2(x_1) & & e_5: & & \dot{x}_2 &= \frac{dx_2}{dt}\\
e_3: & & y &= x_2
\end{aligned}
\label{eq:example_MSO}
\end{equation}
where $g_1$ and $g_2$ are invertible. When each of the equations $e_1$, 
$e_2$, and $e3$, are selected as residual equation, the resulting 
computational graphs are given in \Figure\ref{fig:different_causalities}.
The corresponding computational graph when $e_1$ is used as residual 
equation has derivative causality, when $e_2$ is used it has mixed causality, 
and when $e_3$ is used it has integral causality. 

\begin{figure}[!t]
\centering
\begin{tikzpicture}[scale=0.75]
\node (1y) at (0,0) {\scriptsize $y$};
\node[cvertex] (1e3) at (1,0) {\scriptsize $e_3$};
\node (1x2) at (2,0) {\scriptsize $x_2$};
\node[cvertex] (1e5) at (3,0) {\scriptsize $e_5$};
\node (1dx2) at (4,0) {\scriptsize $\dot{x}_2$};
\node[cvertex] (1e2) at (5,0) {\scriptsize $e_2$};
\node (1x1) at (6,0) {\scriptsize $x_1$};
\node[cvertex] (1e4) at (7,0) {\scriptsize $e_4$};
\node (1dx1) at (8.5,0.5) {\scriptsize $\dot{x}_1$};
\node (1u) at (8.5,-0.5) {\scriptsize $u$};
\node[cvertex] (1e1) at (10,0) {\scriptsize $e_1$};
\node (1r) at (11,0) {\scriptsize $r_1$};
\draw[edge] (1y) to[out=0,in=180] (1e3);
\draw[edge] (1e3) to[out=0,in=180] (1x2);
\draw[edge] (1x2) to[out=0,in=180] (1e5);
\draw[edge] (1e5) to[out=0,in=180] (1dx2);
\draw[edge] (1dx2) to[out=0,in=180] (1e2);
\draw[edge] (1e2) to[out=0,in=180] (1x1);
\draw[edge] (1x1) to[out=0,in=180] (1e4);
\draw[edge] (1e4) to[out=0,in=180] (1dx1);
\draw[edge] (1dx1) to[out=0,in=180] (1e1);
\draw[edge] (1u) to[out=0,in=180] (1e1);
\draw[edge] (1e1) to[out=0,in=180] (1r);

\newcommand\rtwoy{-2.5}
\newcommand\rtwox{2.5}
\node (2y) at (0+\rtwox,0.5+\rtwoy) {\scriptsize $y$};
\node[cvertex] (2e3) at (1+\rtwox,0.5+\rtwoy) {\scriptsize $e_3$};
\node (2x2) at (2+\rtwox,0.5+\rtwoy) {\scriptsize $x_2$};
\node[cvertex] (2e5) at (3+\rtwox,0.5+\rtwoy) {\scriptsize $e_5$};
\node (2dx2) at (4+\rtwox,0.5+\rtwoy) {\scriptsize $\dot{x}_2$};
\node[cvertex] (2e2) at (5.5+\rtwox,0+\rtwoy) {\scriptsize $e_2$};
\node (2u) at (0+\rtwox,-0.5+\rtwoy) {\scriptsize $u$};
\node[cvertex] (2e1) at (1+\rtwox,-0.5+\rtwoy) {\scriptsize $e_1$};
\node (2dx1) at (2+\rtwox,-0.5+\rtwoy) {\scriptsize $\dot{x}_1$};
\node[cvertex] (2e4) at (3+\rtwox,-0.5+\rtwoy) {\scriptsize $e_4$};
\node (2x1) at (4+\rtwox,-0.5+\rtwoy) {\scriptsize $x_1$};

\node (2r) at (6.5+\rtwox,0+\rtwoy) {\scriptsize $r_2$};
\draw[edge] (2y) to[out=0,in=180] (2e3);
\draw[edge] (2e3) to[out=0,in=180] (2x2);
\draw[edge] (2x2) to[out=0,in=180] (2e5);
\draw[edge] (2e5) to[out=0,in=180] (2dx2);
\draw[edge] (2dx2) to[out=0,in=180] (2e2);
\draw[edge] (2e2) to[out=0,in=180] (2r);
\draw[edge] (2u) to[out=0,in=180] (2e1);
\draw[edge] (2e1) to[out=0,in=180] (2dx1);
\draw[edge] (2dx1) to[out=0,in=180] (2e4);
\draw[edge] (2e4) to[out=0,in=180] (2x1);
\draw[edge] (2x1) to[out=0,in=180] (2e2);

\newcommand\rthreey{-5}
\newcommand\rthreex{0}
\node (3u) at (0+\rthreex,0+\rthreey) {\scriptsize $u$};
\node[cvertex] (3e1) at (1+\rthreex,0+\rthreey) {\scriptsize $e_1$};
\node (3dx1) at (2+\rthreex,0+\rthreey) {\scriptsize $\dot{x}_1$};
\node[cvertex] (3e4) at (3+\rthreex,0+\rthreey) {\scriptsize $e_4$};
\node (3x1) at (4+\rthreex,0+\rthreey) {\scriptsize $x_1$};
\node[cvertex] (3e2) at (5+\rthreex,0+\rthreey) {\scriptsize $e_2$};
\node (3dx2) at (6+\rthreex,0+\rthreey) {\scriptsize $\dot{x}_2$};
\node[cvertex] (3e5) at (7+\rthreex,0+\rthreey) {\scriptsize $e_5$};
\node (3x2) at (8.5+\rthreex,0.5+\rthreey) {\scriptsize $x_2$};
\node (3y) at (8.5+\rthreex,-0.5+\rthreey) {\scriptsize $y$};
\node[cvertex] (3e3) at (10+\rthreex,0+\rthreey) {\scriptsize $e_3$};
\node (3r) at (11+\rthreex,0+\rthreey) {\scriptsize $r_3$};
\draw[edge] (3u) to[out=0,in=180] (3e1);
\draw[edge] (3e1) to[out=0,in=180] (3dx1);
\draw[edge] (3dx1) to[out=0,in=180] (3e4);
\draw[edge] (3e4) to[out=0,in=180] (3x1);
\draw[edge] (3x1) to[out=0,in=180] (3e2);
\draw[edge] (3e2) to[out=0,in=180] (3dx2);
\draw[edge] (3dx2) to[out=0,in=180] (3e5);
\draw[edge] (3e5) to[out=0,in=180] (3x2);
\draw[edge] (3x2) to[out=0,in=180] (3e3);
\draw[edge] (3y) to[out=0,in=180] (3e3);
\draw[edge] (3e3) to[out=0,in=180] (3r);
\end{tikzpicture}
\caption{An illustration of three different computational graphs with different causalities using the same MSO set \eqref{eq:example_MSO} but different residual equations.}
\label{fig:different_causalities}
\end{figure}
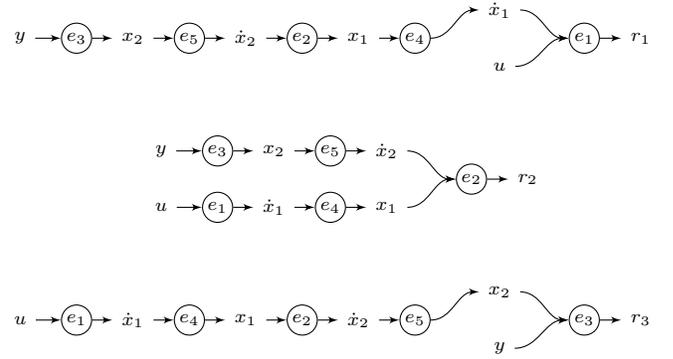
\end{example}

The structure of the computational graph illustrate the structural relationships 
between input and output variables and state variables even though the 
analytical relationships are unknown. This can give useful information to include 
physical insights in the model structure of a recurrent neural network.   
\section{Design of RNN Using Structural Models}

Discrete-time non-linear state-space models can be modeled using RNN. 
Here, computational graphs are derived from MSO sets with integral 
causality. These MSO sets have a DAE index equal to zero or one 
which means that they can be written in state-space form \cite{ascher1998computer}, i.e.
\begin{equation}
\begin{aligned}
\dot{x} &= \bar{g}(x,u) \\
r &= y - h(x,u)
\end{aligned}
\label{eq:tcss}
\end{equation}
where $x$ are state variables, $u$ are known inputs, $y$ is the signal to predict, and $\bar{g} = \left(g_1, g_2, \ldots \right)^T$. Note that $u$ could include both actuator and sensor signals depending on the computational graph. 
The arguments to each function $g_i(x,u): \mathcal{R}^{|x|+|u|} \rightarrow \mathcal{R}$ 
are determined by backtracking from each corresponding state derivative $\dot{x}_i$ in the 
computational graph until a state variable $x$ or an input signal $u$ is found which give 
the arguments to each function $g_i(x, u)$. Similarly, the arguments to the 
function $h(x,u): \mathcal{R}^{|x|+|u|} \rightarrow \mathcal{R}$ are determined by 
back-tracking from the residual equation in the computation graph until a state variable 
or and input signal is found. 

When the arguments have been found, the time-continuous state-space model 
\eqref{eq:tcss} can be formulated in discrete-time, using Euler forward, as 
\begin{equation}
\begin{aligned}
x_{t+1} &= x_{t} + T \, \bar{g}(x_{t},u_{t}) \\
r_{t} &= y_{t} - h(x_t,u_t)
\end{aligned}
\label{eq:tdss}
\end{equation}
where $T$ is the sampling time. 

Once the structure of the discrete-time state space model is determined, an RNN is 
generated with the same structure as the model \eqref{eq:tdss} \cite{pulido2004possible}.
The non-linear functions $g_i$ and $h$ are modeled using a general fully-connected neural network 
structure with a scalar output and the input vector corresponds to the arguments 
in \eqref{eq:tdss} derived from the computational graph, similar to the neural network 
shown in \Figure\ref{fig:ann}. 
\section{Internal Combustion Engine Case Study}

\begin{figure}
\centering
  \begin{tikzpicture}
    \node[anchor=south west,inner sep=0] (image) at (0,0) {
      \resizebox{0.9\columnwidth}{!}{\includegraphics{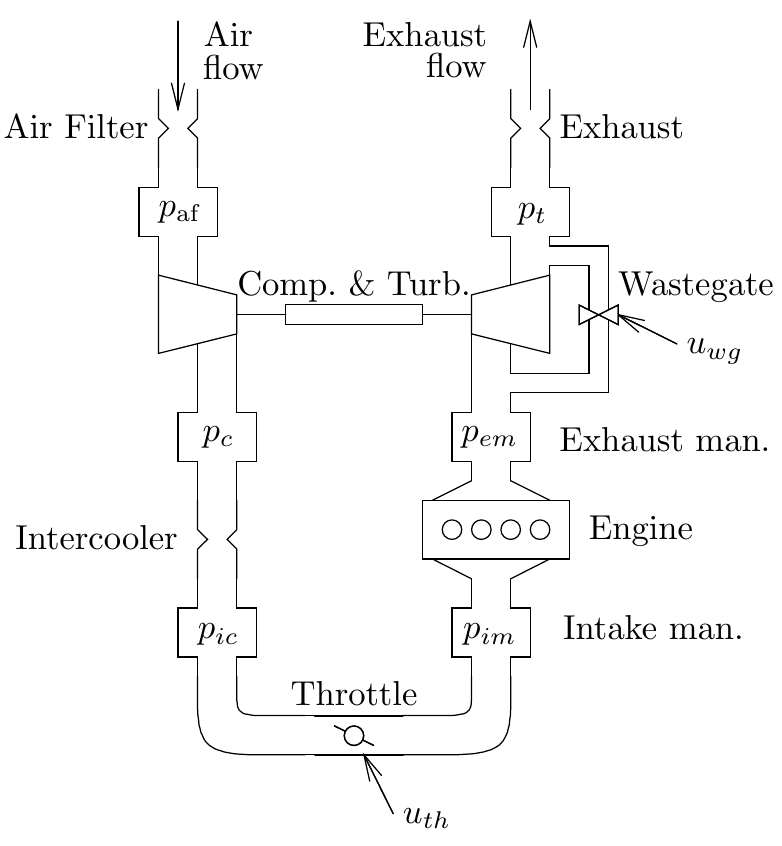}}
    };
    \begin{scope}[x={(image.south east)},y={(image.north west)}]
    \fill[white!0!white] (0.44,0.10825) rectangle (0.48,0.11);
    \fill[white!0!white] (0.7917,0.5) rectangle (1.0,0.64);
    \fill[white!0!white] (0.4,0) rectangle (0.6,0.1068);
    \draw[->] (0.22, 0.22) -- (0.17, 0.21) node[left]{\small $y_{pic}$};
    \draw[->] (0.22, 0.26) -- (0.17, 0.27) node[left]{\small $y_{Tic}$};
    \draw[->] (0.68, 0.21) -- (0.73, 0.20) node[right]{\small $y_{pim}$};
    \draw[->] (0.19, 0.80) -- (0.14, 0.79) node[left]{\small $y_{waf}$};
    \draw[->] (0.53, 0.36) -- (0.48, 0.35) node[left]{\small $y_{\omega}$};
    \draw[->] (0.46, 0.10) -- (0.47, 0.06) node[right]{\small $y_{xpos}$};
    \draw[] (0.00, 0.58) node[right]{\small $y_{pamb}$};
    \draw[] (0.00, 0.52) node[right]{\small $y_{Tamb}$};
    \draw[<-] (0.7917,0.62) -- (0.8417,0.61) node[right]{\small $u_{wg}$};
    \draw[<-] (0.53, 0.39) -- (0.48, 0.41) node[left]{\small $u_{mf}$};
    \end{scope}
  \end{tikzpicture}
  \caption{A schematic of the model of the air flow through the model. This figure is 
  used with permission from \cite{eriksson2002control}.}
  \label{fig:engine_model}
\end{figure}
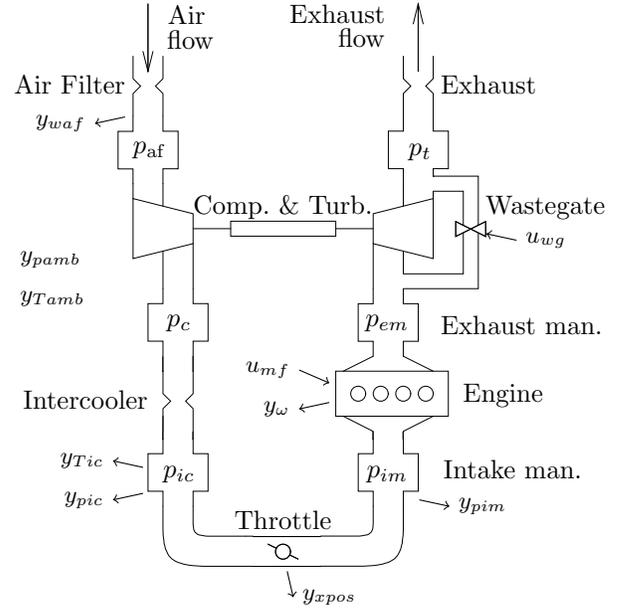

As a case study, the air path through an internal combustion engine is considered 
which is illustrated in \Figure\ref{fig:engine_model}. Both nominal and faulty data 
have been collected from a test bench during transient operation. The engine is 
an interesting case study since the system is dynamic and non-linear and it is used for 
both transients and stationary operating conditions. Another complicating aspect 
from a fault diagnosis perspective is the coupling between the intake and exhaust 
flow through the turbine and compressor. This results in that a fault somewhere in 
the system, including sensor faults, will not have an isolated impact in that component 
but is likely to affect the behavior, and thus the sensor outputs, in other parts of the 
system as well.  

\subsection{Model}

The structural model used in this case study is based on a mathematical mean 
value engine model that has been used in previous works for model-based residual 
generation, see for example \cite{jung2018combining,ng2020design}. The mathematical 
model structure is similar to the model described in \cite{eriksson2007modeling}, which is 
based on six control volumes and mass and energy flows given by restrictions, see 
\Figure\ref{fig:engine_model}. 

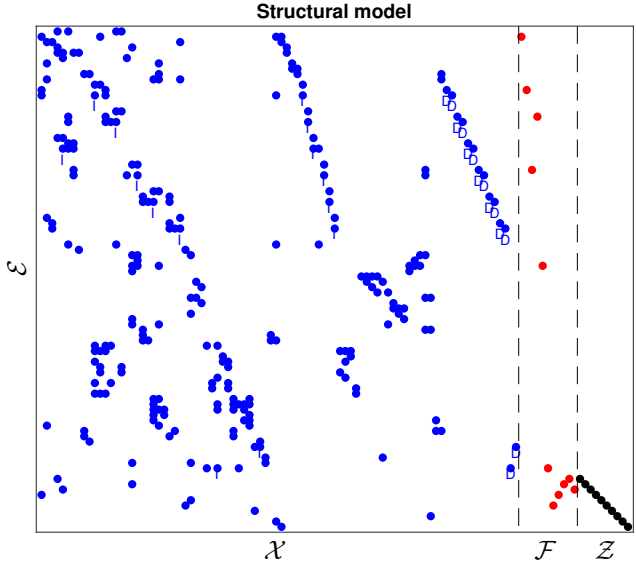
\begin{figure}
    \centering
     \begin{tikzpicture}
     \node[anchor=south west,inner sep=0] (image) at (0,0) {
      \resizebox{0.9\columnwidth}{!}{\includegraphics{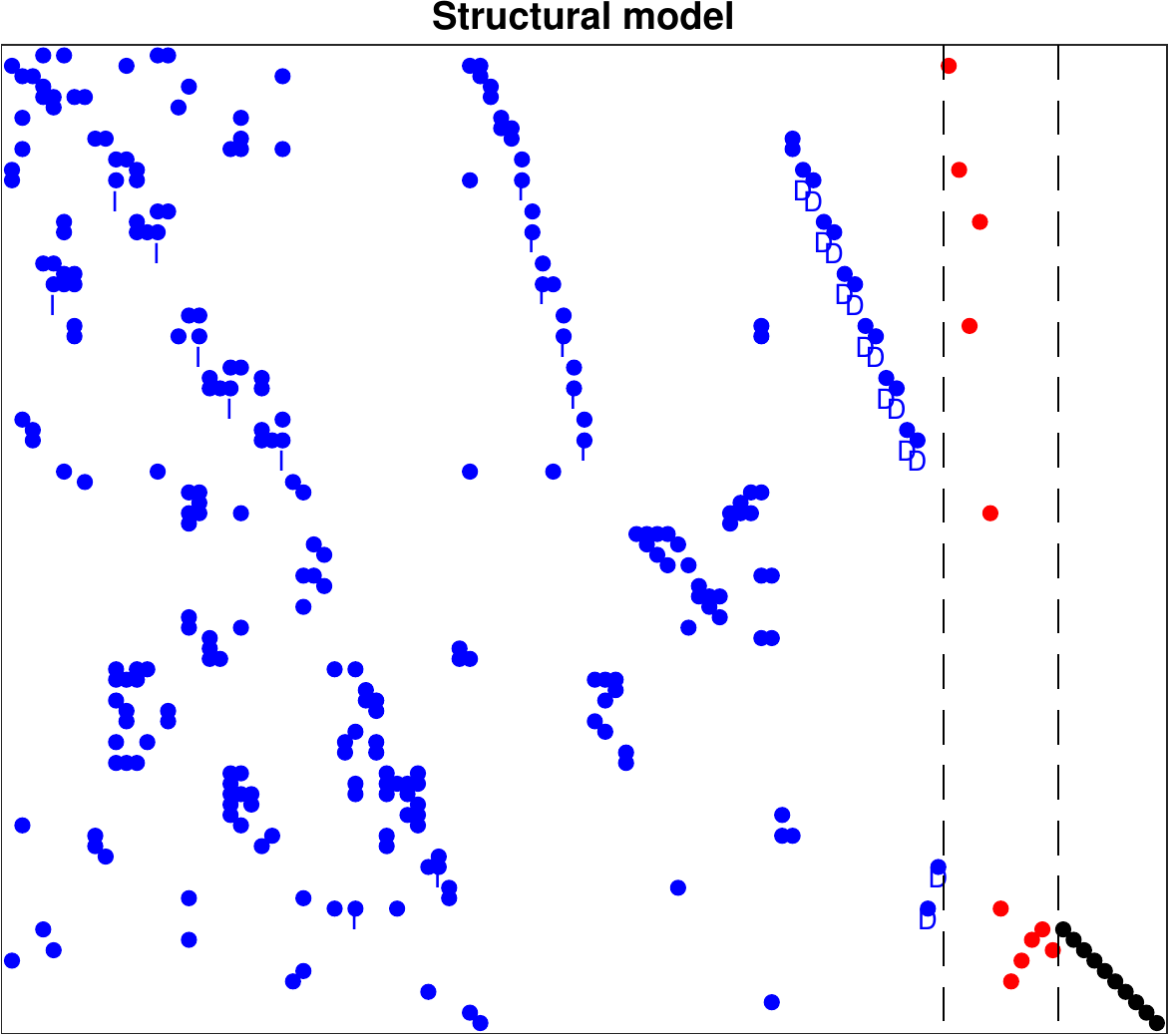}}
    };
    \begin{scope}[x={(image.south east)},y={(image.north west)}]
    \draw (-0.03, 0.50) node[]{\rotatebox{90}{$\mathcal{E}$}};
    \draw (0.4, -0.03) node[]{$\mathcal{X}$};
    \draw (0.85, -0.03) node[]{$\mathcal{F}$};
    \draw (0.95, -0.03) node[]{$\mathcal{Z}$};
    \end{scope}
    \end{tikzpicture}
    \caption{A structural model of the internal combustion engine. }
    \label{fig:structural_model}
\end{figure}

A structural representation of the engine model, used in this work, is shown 
in \Figure\ref{fig:structural_model} where a mark in position $(i,j)$ denotes that 
the variable $x_j$ is included in equation $e_i$. The model variables 
are organized in unknown variables $\mathcal{X}$ (including state variables), known variables $\mathcal{Z}$, 
including known actuators and sensor outputs, and fault signals $\mathcal{F}$. To state the 
relation between state variables and their derivatives in the structural model, 
state variables are marked as I and their derivatives as D in the figure 
\cite{frisk2017toolbox}. The structural model has 94 equations, 90 unknown 
variables, including 14 state variables and their derivatives, 11 fault variables, 
and 10 known variables.
   
\begin{figure}
    \centering
    \includegraphics[width=1.0\columnwidth]{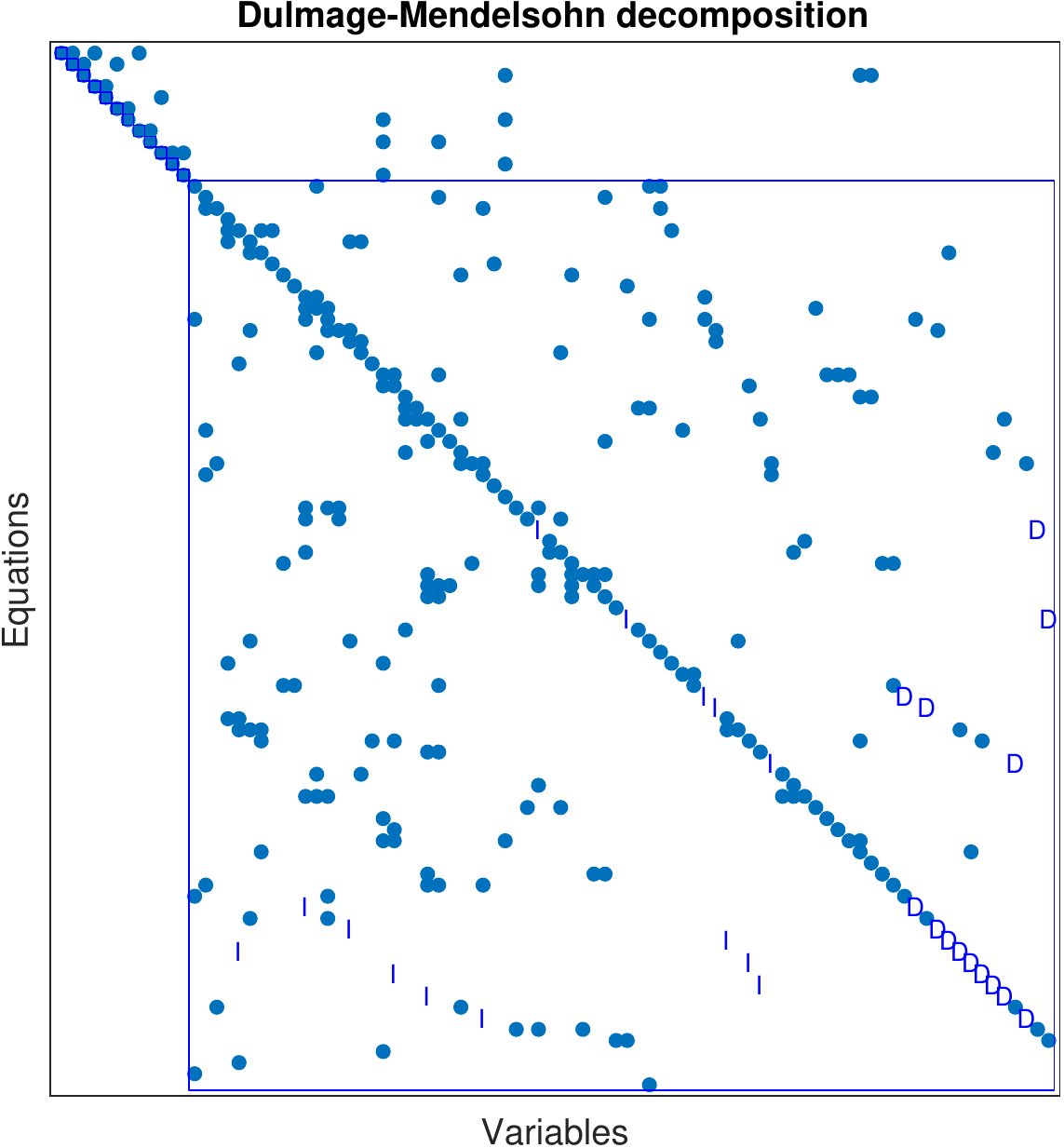}
    \caption{Dulmage-Mendelsohn decomposition of engine model. The blue square represents the over-determined part of the model. State variables and their derivatives are denoted I and D in the matrix.}
    \label{fig:dm_engine}
\end{figure}

A Dulmage-Mendelsohn decomposition of the structural model in 
\Figure\ref{fig:engine_model} is shown in \Figure\ref{fig:dm_engine}.
The over-determine part is shown in the blue square. The variables 
located in the upper left part of the figure belong to the exactly 
determined part of the model.
  
\subsection{Data}

\begin{figure}
    \centering
     \begin{tikzpicture}
     \node[anchor=south west,inner sep=0] (image) at (0,0) {
      \resizebox{1\columnwidth}{!}{\includegraphics{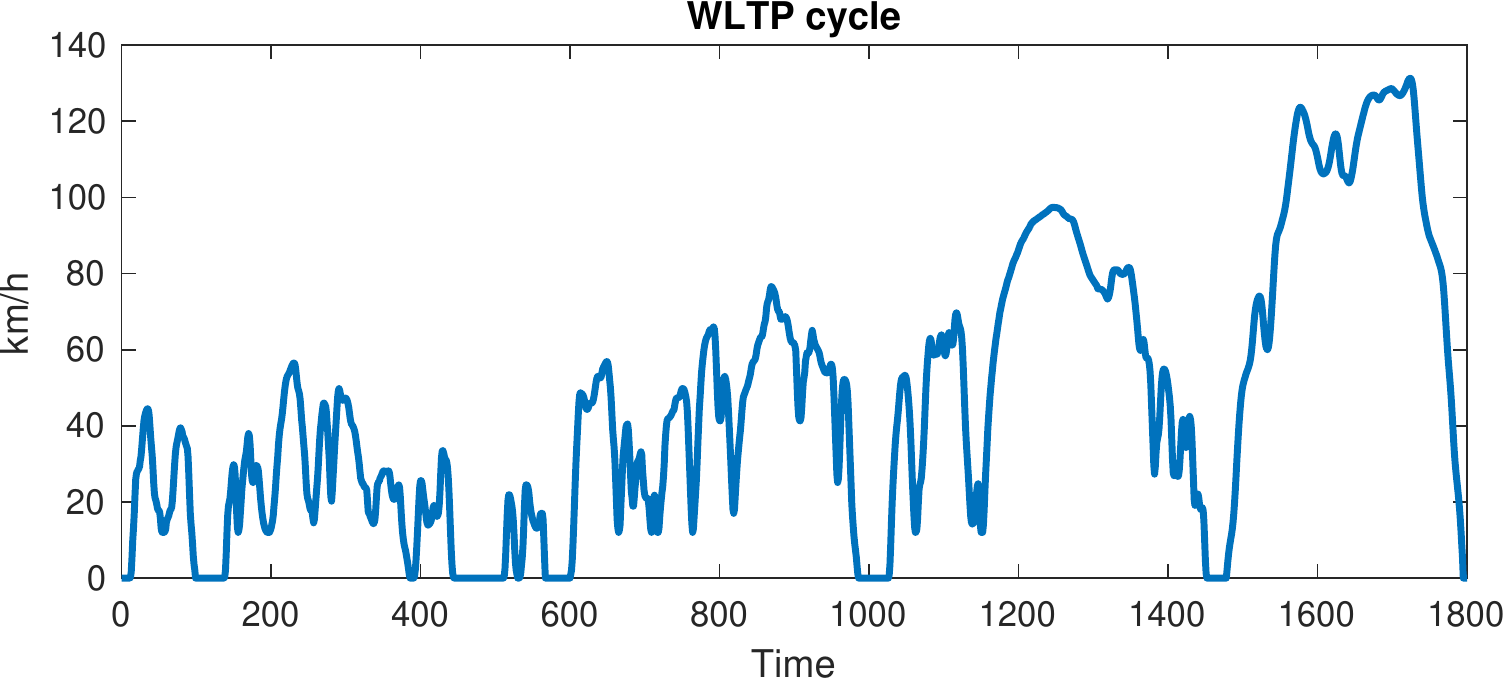}}
    };
    \begin{scope}[x={(image.south east)},y={(image.north west)}]
    \end{scope}
    \end{tikzpicture}
    \caption{Training data have been collected when the engine has been operated to follow the WLTP cycle. }
    \label{fig:wltp}
\end{figure}

Operational data for training and validation have been collected from an 
engine test bench during transient operation. To cover a large range 
of operating conditions, data are collected from the engine when it follows 
the Worldwide Harmonized Light Vehicles Test Procedure (WLTP) cycle, 
see \Figure\ref{fig:wltp}.

\begin{figure}
    \centering
    \includegraphics[width=1\columnwidth]{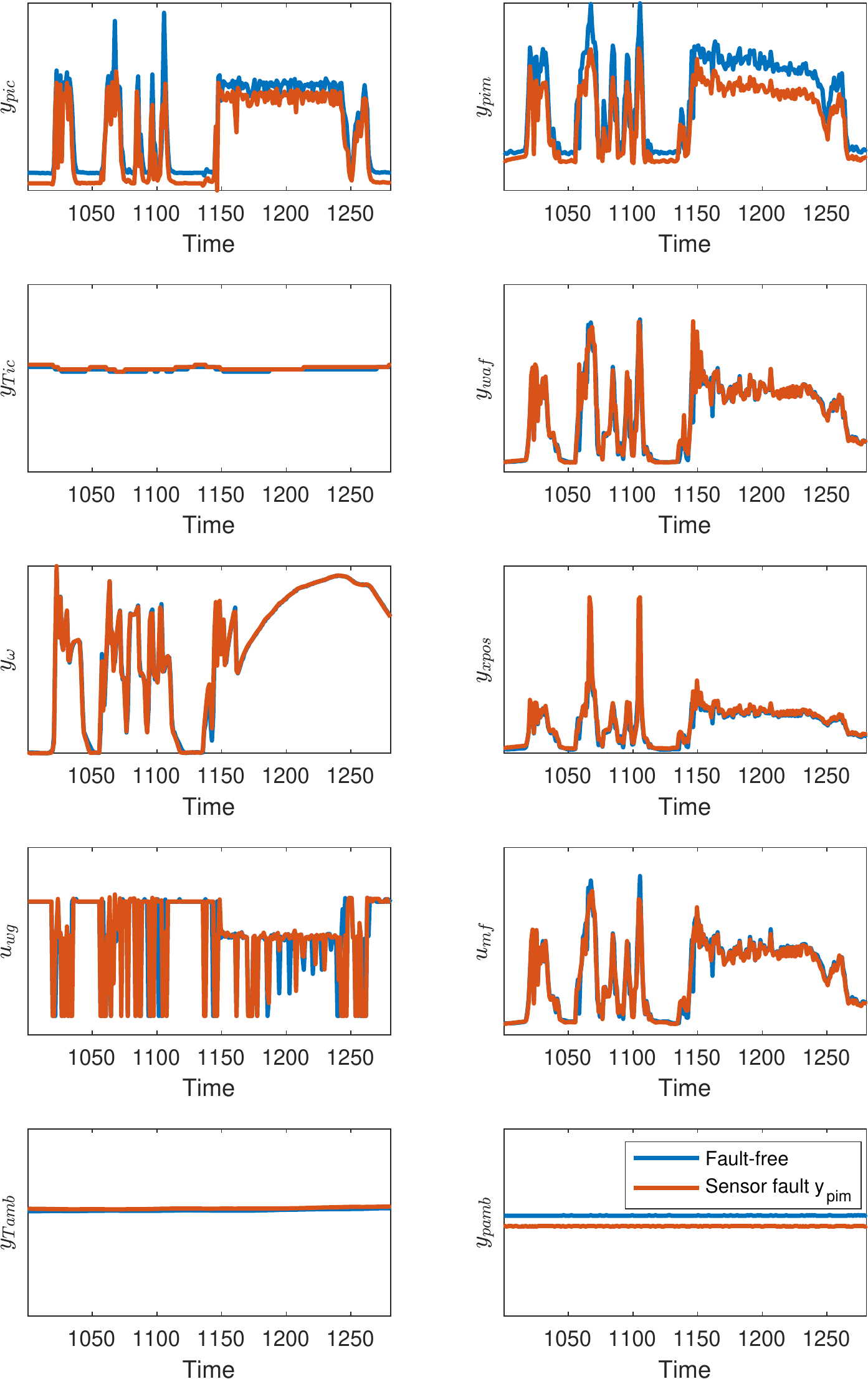}
    \caption{Example of engine data from fault-free and faulty operation when there is a fault in sensor $y_{pim}$. Note that the sensor fault affects the system behavior and is visible in other signals as well, e.g. $y_{pic}$.}
    \label{fig:engine_signals}
\end{figure}

The set of available sensors in the engine corresponds to a standard setup 
used in a conventional car. A list of which sensor data are collected in each 
data set is summarized in Table~\ref{tab:signals} and some examples of 
collected data from nominal and faulty operation (fault in sensor $y_{pim}$) 
are shown in \Figure\ref{fig:engine_signals}. Data are downsampled from 
1 kHz to 20 Hz to reduce the computational complexity while still capturing the dynamics 
of the engine. Each signal is normalized such that nominal signal output is in 
the range of about [0, 1].

\begin{table}
\caption{A summary of actuator and sensor signals collected from the engine.}
\label{tab:signals}
\centering
\begin{tabular}{p{0.7cm} p{7.3cm}}
\hline
Signal & Description \\
\hline
$u_{wg}$ & Wastegate reference signal \\
$u_{mf}$ &  Estimated total fuel mass flow in all cylinders \\
$y_{pic}$ & Measured pressure at intercooler \\
$y_{pim}$ & Measured pressure at intake manifold \\  
$y_{Tic}$ & Measured temperature at intercooler \\  
$y_{waf}$ & Measured air mass flow through air filter \\  
$y_{\omega}$ & Measured engine speed \\
$y_{xpos}$ & Measured throttle angle \\
$y_{pamb}$ & Measured ambient pressure \\
$y_{Tamb}$ & Measured ambient temperature \\
\hline
\end{tabular}
\end{table}

Experimental data have been collected from different sensor 
fault scenarios and leakages, see Table~\ref{tab:faults}. Multiplicative 
sensor faults of different magnitudes have been injected by modifying 
the corresponding signal directly in the engine control unit during 
operation. This is illustrated in \Figure\ref{fig:engine_signals} comparing 
engine sensor data from nominal case and scenario with fault in sensor 
measuring intake manifold pressure. The sensor fault has impact on the 
general system operation and is affecting other measured states as well.

\begin{table}
\caption{A summary of fault scenarios collected from engine test bench.}
\label{tab:faults}
\centering
\begin{tabular}{p{0.7cm} p{7.3cm}}
\hline
Fault & Description \\
\hline
$f^{Wth}$ & Leakage before throttle \\
$f^{ypic}$ & Intermittent fault in sensor measuring pressure at intercooler \\  
$f^{ypim}$ & Intermittent fault in sensor measuring intake manifold pressure \\  
$f^{ywaf}$ & Intermittent fault in sensor measuring air flow through air filter \\  
\hline
\end{tabular}
\end{table}
\section{Experiments}

A set of grey-box RNN is generated based on MSO sets derived from 
the structural model. The RNN model structures are designed using 
computational graphs with integral causality and where the residual equation 
is a sensor equation, i.e. $r_t = y_t - \hat{y}_t$. The steps of the 
design procedure are summarized in \Figure\ref{fig:procedure}.
Then, the generated grey-box RNN are evaluated as residual generators
using collected data from the engine test bench. With a slight abuse of notation, the 
MSO set index MSO$_i$ will be used in the text when referring to the 
generated grey-box RNN and residual generator.    

\tikzstyle{box} = [rectangle, minimum width= 0.5cm, minimum height= 4.8cm, text width=0.8cm, text centered, draw=black]
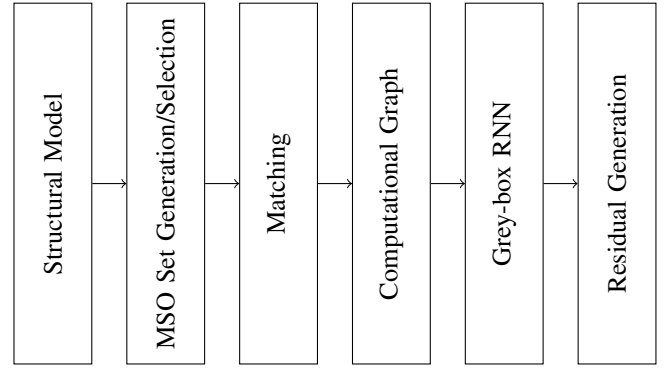
\begin{figure}
\centering
\begin{tikzpicture}[node distance = 1.5cm]
\node (sm) [box] {\rotatebox{90}{Structural Model}};
\node (mso) [box, right of=sm] {\rotatebox{90}{MSO Set Generation/Selection}};
\node (ma) [box, right of=mso] {\rotatebox{90}{Matching}};
\node (cg) [box, right of=ma] {\rotatebox{90}{Computational Graph}};
\node (rnn) [box, right of=cg] {\rotatebox{90}{Grey-box RNN}};
\node (res) [box, right of=rnn] {\rotatebox{90}{Residual Generation}};
\draw[->] (sm) -- (mso); 
\draw[->] (mso) -- (ma); 
\draw[->] (ma) -- (cg); 
\draw[->] (cg) -- (rnn); 
\draw[->] (rnn) -- (res); 
\end{tikzpicture}
\caption{The design procedure of generating residual generators using grey-box RNN from a structural model.}
\label{fig:procedure}
\end{figure}

\subsection{Residual Generation}

First, a set of MSO sets is computed based on the structural model using the 
Fault Diagnosis Toolbox \cite{frisk2017toolbox} in Matlab. A causality analysis 
is performed on each of the 144 candidate MSO sets to identify which MSO 
sets that can be used to generate computational graphs with integral causality, 
resulting in 17 sets. Out of these 17 MSO sets it is possible to generate 21 
different computational graphs with integral causality. The grey-box RNN and 
resulting residual generators will be referred to in the text based on which of the 144 
original MSO sets they are generated from.  

Only the computational graphs where one of the sensors $y_{pim}$, $y_{pic}$, or 
$y_{waf}$ are used as residual equation, are used to generate grey-box recurrent 
neural networks to be used in this case study. The other candidates have residual 
equations based on sensors measuring slowly varying states, such as temperatures, 
as shown in \Figure\ref{fig:engine_signals}. These sensor signals do not show 
sufficient excitation in training data and are, therefore, not used as residual equations. 

\begin{figure}
    \centering
    \includegraphics[width=1.0\columnwidth]{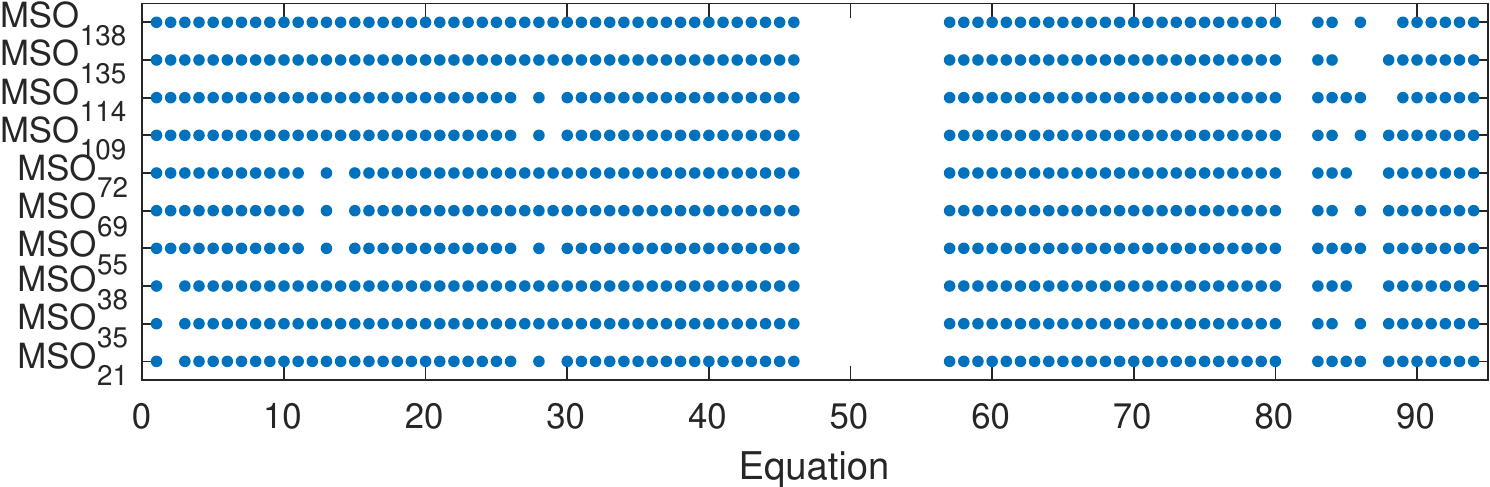}
    \caption{The model support of each evaluated MSO set.}
    \label{fig:mtes_support}
\end{figure}

The model support of each MSO set that is used for generating computational graphs in 
this study, is shown in \Figure\ref{fig:mtes_support}. A mark in position $(i,j)$ means that 
equation $e_j$ is included in MSO set $i$. It is visible that the MSO sets share most 
of the equations in this case study. This can be explained by that the engine has few 
sensors and is strongly coupled meaning that a large part of the model is needed to 
achieve redundancy. Note that equations $e_{47}$-$e_{56}$ are not included in any 
MSO set. These equations corresponds to the equations not included in the 
over-determined part of the model $\mathcal{M}^+$ in \Figure\ref{fig:dm_engine}.

\begin{figure}
    \centering
    \includegraphics[width=1.0\columnwidth]{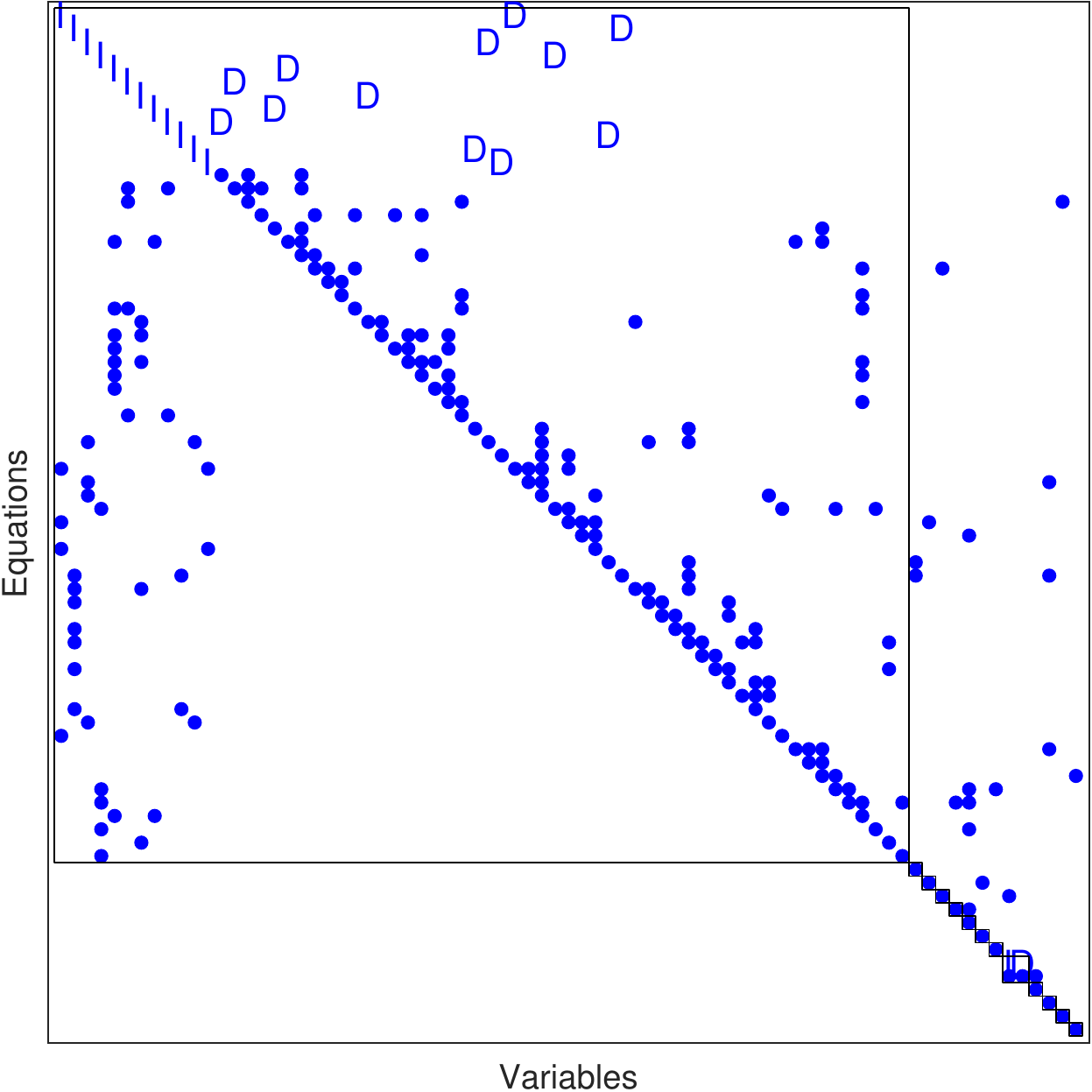}
    \caption{Matching of MSO$_{27}$ used to derive computational graph. Starting from 
    the lowest row, each unknown variable on the diagonal can be computed based on 
    previously computed variables are states.}
    \label{fig:matching}
\end{figure}

The generated computational graphs are then used to automatically generate a set 
of grey-box RNN. An incidence matrix used to derive the matching of MSO$_{27}$ 
is shown in \Figure\ref{fig:matching} going down-to-up. The incidence matrix 
represents a bi-partite graph and the matching is interpreted as each variable 
in the diagonal can be computed based on the variables marked on the same 
row. Rows with I on the diagonal denotes computation of state variables by 
integrating their derivatives. The result is a computational graph representing 
the state-space model which only depends on known variables and state 
variables \eqref{eq:tcss}.   

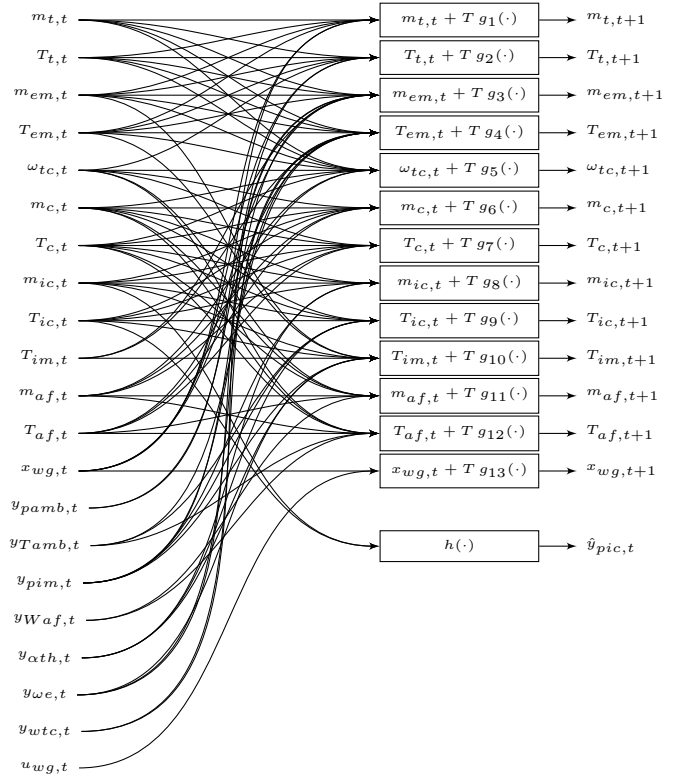
\begin{figure}[!t]
\centering
\begin{tikzpicture}
\node[vertex] (m_t) at (0,0) {\tiny $m_{t,t} + T \,g_1(\cdot)$};
\node[vertex] (T_t) at (0,-0.5) {\tiny $T_{t,t} + T \,g_2(\cdot)$};
\node[vertex] (m_em) at (0,-1) {\tiny $m_{em,t} + T \,g_3(\cdot)$};
\node[vertex] (T_em) at (0,-1.5) {\tiny $T_{em,t} + T \,g_4(\cdot)$};
\node[vertex] (omega_tc) at (0,-2) {\tiny $\omega_{tc,t} + T \,g_5(\cdot)$};
\node[vertex] (m_c) at (0,-2.5) {\tiny $m_{c,t} + T \,g_6(\cdot)$};
\node[vertex] (T_c) at (0,-3) {\tiny $T_{c,t} + T \,g_7(\cdot)$};
\node[vertex] (m_ic) at (0,-3.5) {\tiny $m_{ic,t} + T \,g_8(\cdot)$};
\node[vertex] (T_ic) at (0,-4) {\tiny $T_{ic,t} + T \,g_9(\cdot)$};
\node[vertex] (T_im) at (0,-4.5) {\tiny $T_{im,t} + T \,g_{10}(\cdot)$};
\node[vertex] (m_af) at (0,-5) {\tiny $m_{af,t} + T \,g_{11}(\cdot)$};
\node[vertex] (T_af) at (0,-5.5) {\tiny $T_{af,t} + T \,g_{12}(\cdot)$};
\node[vertex] (wg_pos) at (0,-6) {\tiny $x_{wg,t} + T \,g_{13}(\cdot)$};
\node[vertex] (h) at (0,-7) {\tiny $h(\cdot)$};

\node[left = 4cm of m_t] (m_t_t) {\tiny $m_{t,t}$};
\node[left = 4cm of T_t] (T_t_t) {\tiny $T_{t,t}$};
\node[left = 4cm of m_em] (m_em_t) {\tiny $m_{em,t}$};
\node[left = 4cm of T_em] (T_em_t) {\tiny $T_{em,t}$};
\node[left = 4cm of omega_tc] (omega_tc_t) {\tiny $\omega_{tc,t}$};
\node[left = 4cm of m_c] (m_c_t) {\tiny $m_{c,t}$};
\node[left = 4cm of T_c] (T_c_t) {\tiny $T_{c,t}$};
\node[left = 4cm of m_ic] (m_ic_t) {\tiny $m_{ic,t}$};
\node[left = 4cm of T_ic] (T_ic_t) {\tiny $T_{ic,t}$};
\node[left = 4cm of T_im] (T_im_t) {\tiny $T_{im,t}$};
\node[left = 4cm of m_af] (m_af_t) {\tiny $m_{af,t}$};
\node[left = 4cm of T_af] (T_af_t) {\tiny $T_{af,t}$};
\node[left = 4cm of wg_pos] (wg_pos_t) {\tiny $x_{wg,t}$};

\node[below = 0.08cm of wg_pos_t] (y_p_amb) {\tiny $y_{pamb,t}$};
\node[below = 0.08cm of y_p_amb] (y_T_amb) {\tiny $y_{Tamb,t}$};
\node[below = 0.08cm of y_T_amb] (y_p_im) {\tiny $y_{pim,t}$};
\node[below = 0.08cm of y_p_im] (y_W_af) {\tiny $y_{Waf,t}$};
\node[below = 0.08cm of y_W_af] (y_alpha_th) {\tiny $y_{\alpha th,t}$};
\node[below = 0.08cm of y_alpha_th] (y_omega_e) {\tiny $y_{\omega e,t}$};
\node[below = 0.08cm of y_omega_e] (y_wfc) {\tiny $y_{wtc,t}$};
\node[below = 0.08cm of y_wfc] (y_u_wg) {\tiny $u_{wg,t}$};

\node[right = 0.5cm of m_t] (m_t_tp1) {\tiny $m_{t,t+1}$};
\node[right = 0.5cm of T_t] (T_t_tp1) {\tiny $T_{t,t+1}$};
\node[right = 0.5cm of m_em] (m_em_tp1) {\tiny $m_{em,t+1}$};
\node[right = 0.5cm of T_em] (T_em_tp1) {\tiny $T_{em,t+1}$};
\node[right = 0.5cm of omega_tc] (omega_tc_tp1) {\tiny $\omega_{tc,t+1}$};
\node[right = 0.5cm of m_c] (m_c_tp1) {\tiny $m_{c,t+1}$};
\node[right = 0.5cm of T_c] (T_c_tp1) {\tiny $T_{c,t+1}$};
\node[right = 0.5cm of m_ic] (m_ic_tp1) {\tiny $m_{ic,t+1}$};
\node[right = 0.5cm of T_ic] (T_ic_tp1) {\tiny $T_{ic,t+1}$};
\node[right = 0.5cm of T_im] (T_im_tp1) {\tiny $T_{im,t+1}$};
\node[right = 0.5cm of m_af] (m_af_tp1) {\tiny $m_{af,t+1}$};
\node[right = 0.5cm of T_af] (T_af_tp1) {\tiny $T_{af,t+1}$};
\node[right = 0.5cm of wg_pos] (wg_pos_tp1) {\tiny $x_{wg,t+1}$};
\node[right = 0.5cm of h] (yhat) {\tiny $\hat{y}_{pic,t}$};

\draw[edge] (m_t) to[out=0,in=180] (m_t_tp1);
\draw[edge] (T_t) to[out=0,in=180] (T_t_tp1);
\draw[edge] (m_em) to[out=0,in=180] (m_em_tp1);
\draw[edge] (T_em) to[out=0,in=180] (T_em_tp1);
\draw[edge] (omega_tc) to[out=0,in=180] (omega_tc_tp1);
\draw[edge] (m_c) to[out=0,in=180] (m_c_tp1);
\draw[edge] (T_c) to[out=0,in=180] (T_c_tp1);
\draw[edge] (m_ic) to[out=0,in=180] (m_ic_tp1);
\draw[edge] (T_ic) to[out=0,in=180] (T_ic_tp1);
\draw[edge] (T_im) to[out=0,in=180] (T_im_tp1);
\draw[edge] (m_af) to[out=0,in=180] (m_af_tp1);
\draw[edge] (T_af) to[out=0,in=180] (T_af_tp1);
\draw[edge] (wg_pos) to[out=0,in=180] (wg_pos_tp1);
\draw[edge] (h) to[out=0,in=180] (yhat);
\foreach \x in {m_t_t, T_t_t, m_em_t, T_em_t, wg_pos_t, y_p_amb} {%
    \draw[edge] (\x) to[out=0,in=180] (m_t);
}
\foreach \x in {m_t_t, T_t_t, m_em_t, T_em_t, omega_tc_t, wg_pos_t, y_p_amb} {%
    \draw[edge] (\x) to[out=0,in=180] (T_t);
}
\foreach \x in {m_t_t, T_t_t, m_em_t, T_em_t, T_im_t, wg_pos_t, y_p_im, y_omega_e, y_wfc} {%
    \draw[edge] (\x) to[out=0,in=180] (m_em);
}
\foreach \x in {m_t_t, T_t_t, m_em_t, T_em_t, T_im_t, wg_pos_t, y_p_im, y_omega_e, y_wfc, y_T_amb} {%
    \draw[edge] (\x) to[out=0,in=180] (T_em);
}
\foreach \x in {m_t_t, T_t_t, m_em_t, T_em_t, omega_tc_t, m_c_t, T_c_t, m_af_t, T_af_t} {%
    \draw[edge] (\x) to[out=0,in=180] (omega_tc);    
}
\foreach \x in {m_ic_t, T_ic_t, omega_tc_t, m_c_t, T_c_t, m_af_t, T_af_t} {%
    \draw[edge] (\x) to[out=0,in=180] (m_c);    
}
\foreach \x in {m_ic_t, T_ic_t, omega_tc_t, m_c_t, T_c_t, m_af_t, T_af_t} {%
    \draw[edge] (\x) to[out=0,in=180] (T_c);    
}
\foreach \x in {m_c_t, T_c_t, m_ic_t, T_ic_t, y_p_im, y_alpha_th} {%
    \draw[edge] (\x) to[out=0,in=180] (m_ic);    
}  
\foreach \x in {m_em_t, T_em_t, m_ic_t, T_ic_t, T_im_t, y_p_im, y_omega_e, y_alpha_th} {%
    \draw[edge] (\x) to[out=0,in=180] (T_im);    
}  
\foreach \x in {m_c_t, T_c_t, m_ic_t, T_ic_t, y_p_im, y_alpha_th, y_T_amb} {%
    \draw[edge] (\x) to[out=0,in=180] (T_ic);    
}  
\foreach \x in {omega_tc_t, m_c_t, T_c_t, m_af_t, T_af_t, y_W_af} {%
    \draw[edge] (\x) to[out=0,in=180] (m_af);    
}  
\foreach \x in {omega_tc_t, m_c_t, T_c_t, m_af_t, T_af_t, y_W_af, y_T_amb} {%
    \draw[edge] (\x) to[out=0,in=180] (T_af);
}  
\foreach \x in {wg_pos_t, y_u_wg} {%
    \draw[edge] (\x) to[out=0,in=180] (wg_pos);
}
\foreach \x in {T_ic_t, m_ic_t} {
    \draw[edge] (\x) to[out=0,in=180] (h);
}

\end{tikzpicture}
\caption{A schematic of the grey-box RNN estimating the output $\hat{y}_{pic,t}$ based 
on MSO$_{27}$ where each function $g_i(\cdot): \mathbb{R}^{n_i} \rightarrow \mathbb{R}$ 
and $h(\cdot): \mathbb{R}^{n_h} \rightarrow \mathbb{R}$ is modeled using a static neural network.}
\label{fig:rnn}
\end{figure}

The grey-box RNN, generated based on the computational graph derived 
from \Figure\ref{fig:matching}, is shown in \Figure\ref{fig:rnn}. 
The non-linear functions $g_i(\cdot)$ and $h(\cdot)$ are modeled using a three layer 
neural network structure, with 256 nodes in each hidden layer, and a scalar output, 
similar as the one illustrated in 
\Figure\ref{fig:ann}. The dimension of the input layer is determined 
from the computational graph. Different activation functions have been evaluated 
where ReLU gave the overall best results and is therefore used in all grey-box RNN 
generated in this case study.  

\subsection{Training}

\begin{figure}
    \centering
    \includegraphics[width=1\columnwidth]{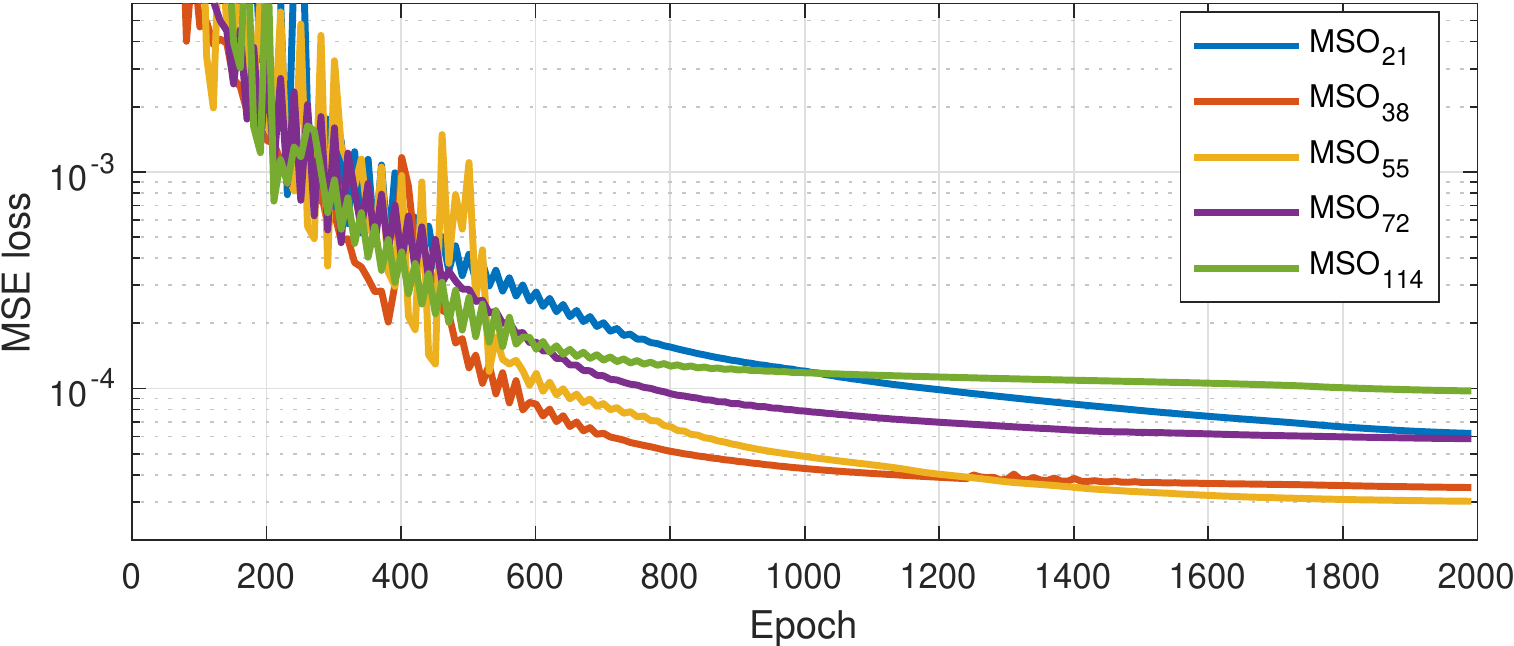}
    \caption{Loss after each epoch when training neural networks predicting $y_{pic}$.}
    \label{fig:fpic_loss}
\end{figure}

\begin{figure}
    \centering
    \includegraphics[width=1\columnwidth]{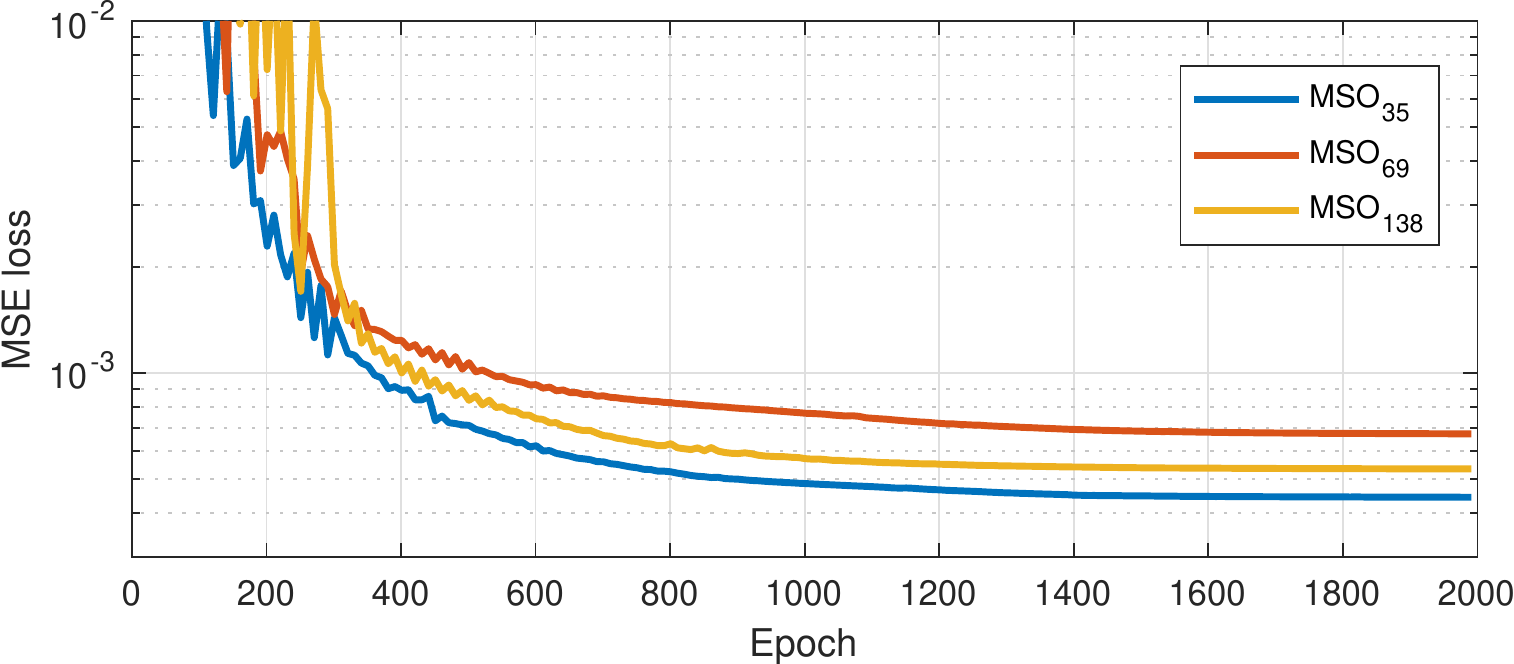}
    \caption{Loss after each epoch when training neural networks predicting $y_{pim}$.}
    \label{fig:fpim_loss}
\end{figure}

Each neural network is trained using only fault-free data where the time series data are 
partitioned into a set of 19 equally sized batches of 600 samples covering different operating 
conditions. Training is done in Python, using PyTorch \cite{paszke2017automatic}, by minimizing the  mean 
square prediction error $\frac{1}{N}\sum_{t=1}^{N} (y_t - \hat{y}_t)^2$. The initial values of the state variables 
are unknown and are set to some reasonable reference value that is used for all batches. 
The optimization algorithm ADAM \cite{diederik2015adam} is run for 2000 epochs with an 
adaptive learning rate starting at $5*10^{-4}$ which is reduced every 10th epoch by $3 \%$. 
The loss after each epoch is shown in \Figure\ref{fig:fpic_loss} for grey-box RNN predicting 
$\hat{y}_{pic}$ and in \Figure\ref{fig:fpim_loss} for grey-box RNN predicting 
$\hat{y}_{pim}$. 

\begin{figure}
    \centering
    \includegraphics[width=1\columnwidth]{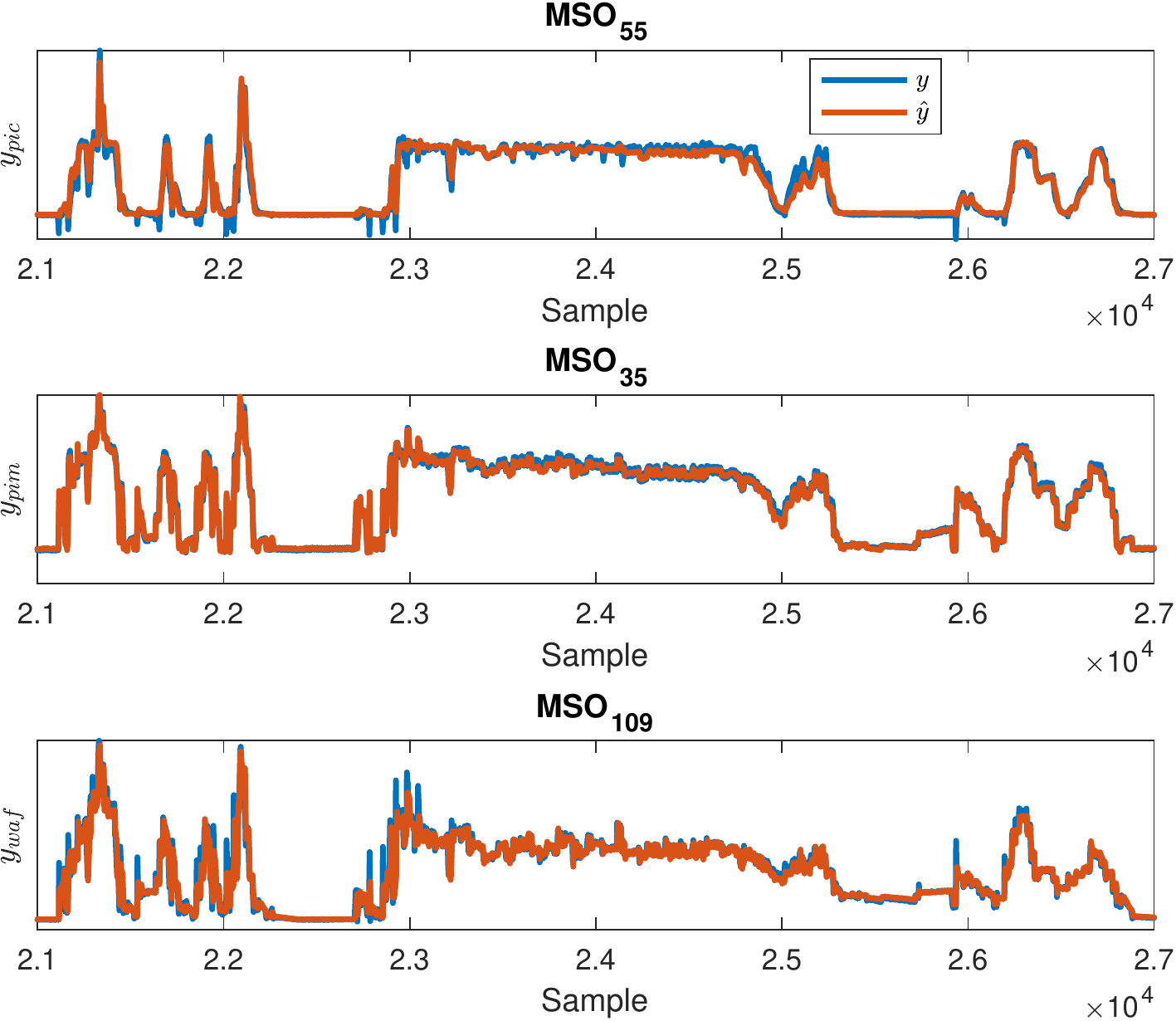}
    \caption{Evaluation of prediction performance of three grey-box RNN.}
    \label{fig:model_predictions}
\end{figure}

Model prediction of three grey-box RNN set are shown in \Figure\ref{fig:model_predictions}, 
each one predicting one of the different sensor outputs. It is visible that the different 
grey-box RNN capture the general dynamic behavior of the different sensor signals. Similar 
prediction performance is achieved for the remaining grey-box RNN. The residual outputs have 
a small bias in the datasets that are likely to be caused by incorrect initial conditions of the state 
variables. If it is assumed that there are no faults when a scenario begins, the residual bias can 
be removed by subtracting the median computed from a short time interval in the beginning of 
the residual output for each dataset.  

\subsection{Fault Diagnosis Performance}

\begin{figure}
    \centering
    \includegraphics[width=1\columnwidth]{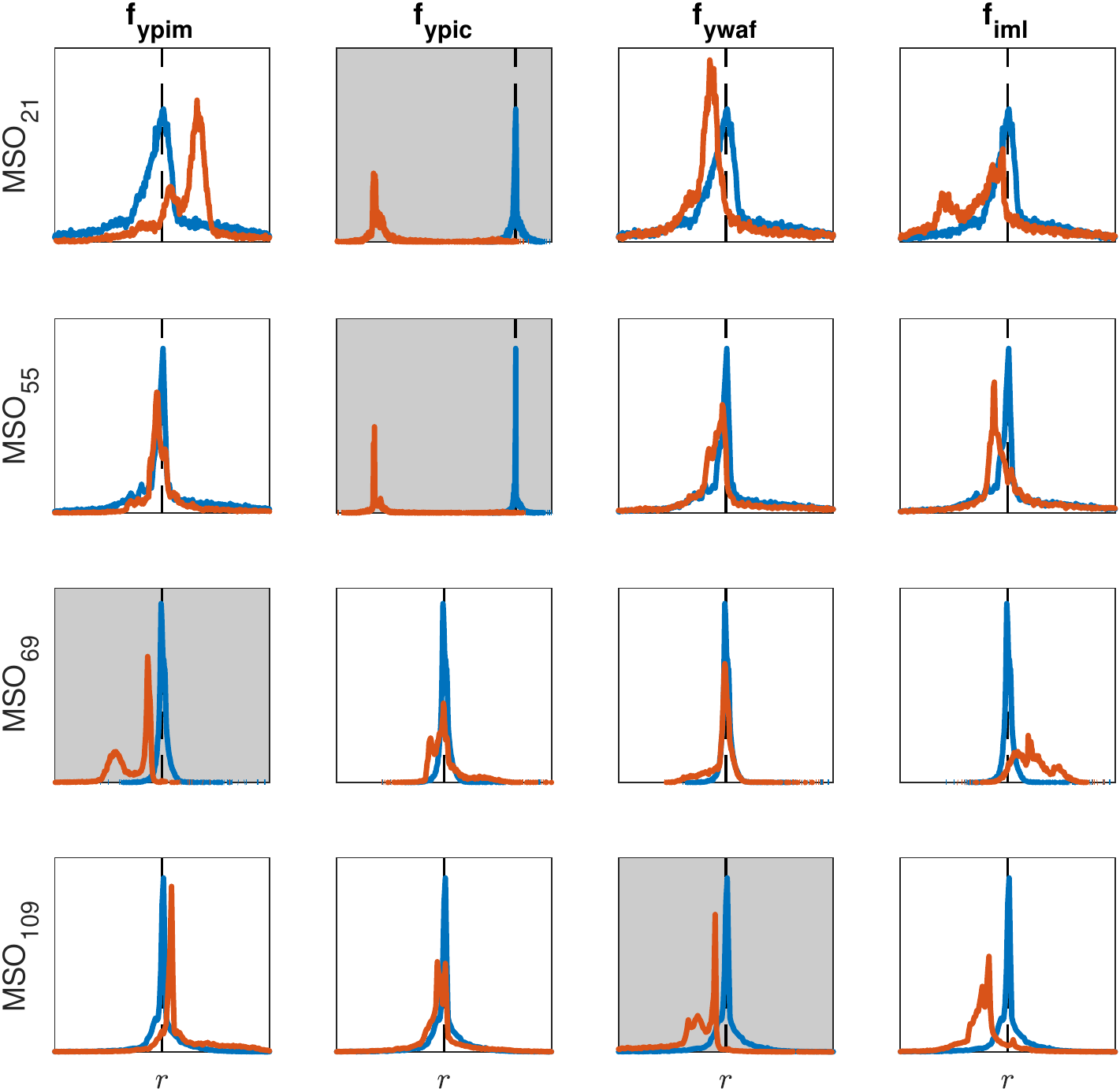}
    \caption{Histograms of residual outputs during different fault scenarios. The blue curve represents 
    the residual output during fault-free case and the red curve when a fault is present. Plots are highlighted 
    when the sensor fault is affecting the sensor signal that predicted by the grey-box RNN used in the 
    residual generator.}
    \label{fig:r_histograms}
\end{figure}

Fault detection performance is evaluated for the set of residual generators using data 
from different fault realizations. Figure~\ref{fig:r_histograms} shows four of the residual 
generators where their outputs are plotted as histograms comparing different fault 
scenarios and nominal operation. The sensor fault scenarios are multiplicative of magnitude 
$f = -20\%$, i.e. $y = (1+f)x$. 
The sensor that is predicted in each residual generator is highlighted in gray in 
\Figure\ref{fig:r_histograms}. Note that faults in predicted sensor signals have a clear impact on 
the residual outputs while detection performance for other faults is varying between different residuals. 
    
\begin{figure}
    \centering
    \includegraphics[width=1\columnwidth]{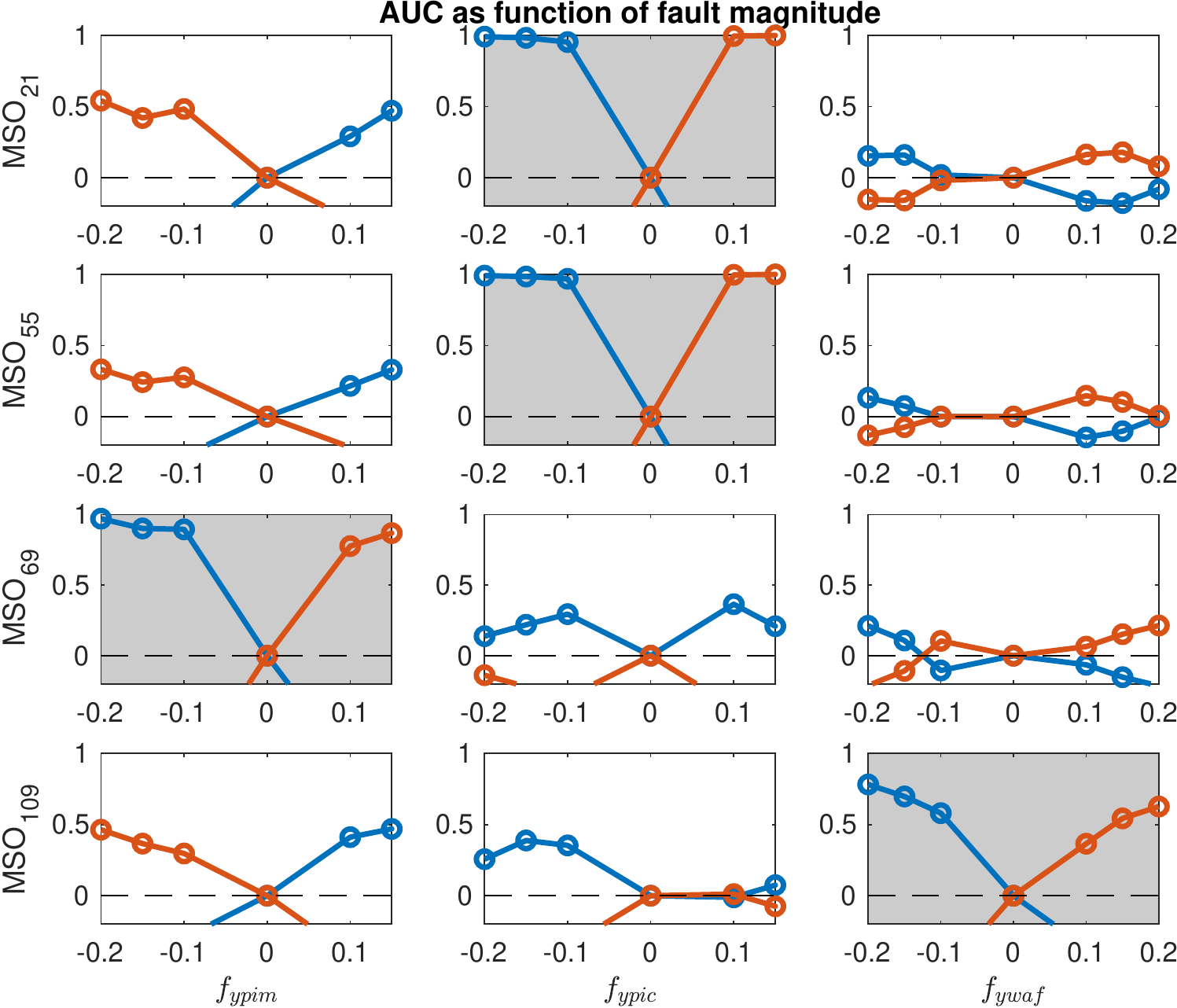}
    \caption{ROC area under curve as function of fault size. The blue curve represents the area under the 
    ROC curve for a test $r > J$ and the red curve a test $-r > J$ when varying the threshold $J$.}
    \label{fig:auc}
\end{figure}

To further analyze detection performance, a Receiver Operating Characteristics (ROC) curve is computed 
for different fault magnitudes. Figure~\ref{fig:auc} shows detection performance for a couple of different of 
residuals by evaluating the area under the curve (AUC) is plotted as function of fault size. Multiplicative 
sensor faults are evaluated for magnitudes in the range [-20\%, 20\%]. Since $\text{AUC} = 0.5$ corresponds 
to that the histograms from nominal and fault case are identical, the plotted AUC curves in \Figure\ref{fig:auc} 
are normalized as $2*(\text{AUC}-0.5)$. Faults affecting the predicted sensor signal in each residual are 
highlighted in gray. It is visible that these faults are easiest to detect by each residual. Some faults have little 
or no impact on the residual outputs, e.g. the residuals based on 
MSO$_{21}$, MSO$_{55}$, and MSO$_{69}$ are not sensitive to fault $f_{ywaf}$ since AUC is almost zero 
for all fault magnitudes.

Even though detection performance varies between the different residual generators, the multi-variate fault 
patterns are still useful for fault classification. A t-distributed Stochastic Nearest Embedding (t-SNE) plot is 
generated to visualize how well the set of residual generators are able to separate data from different faults 
classes. The t-SNE plot is a non-linear projection of high-dimensional data to a lower-dimensional space 
where pair-wise similarities between samples are preserved \cite{maaten2008visualizing}. Residual outputs 
from the four residual generators in \Figure\ref{fig:r_histograms} using data from all different fault classes 
and fault realizations are used as input to the t-SNE algorithm. Data is then down-sampled to improve 
visualization. It is visible in the figure that data from the different faults are not overlapping with the fault-free 
data points, indicating that all faults are detectable. It is also visible that most faults are distinguishable from 
each other except that data from a leakage in the intake manifold $f_{iml}$ are overlapping with data from 
sensor fault $f_{ypim}$ which is reasonable since they are located in the same part of the engine. 

\begin{figure}
    \centering
    \includegraphics[width=1\columnwidth]{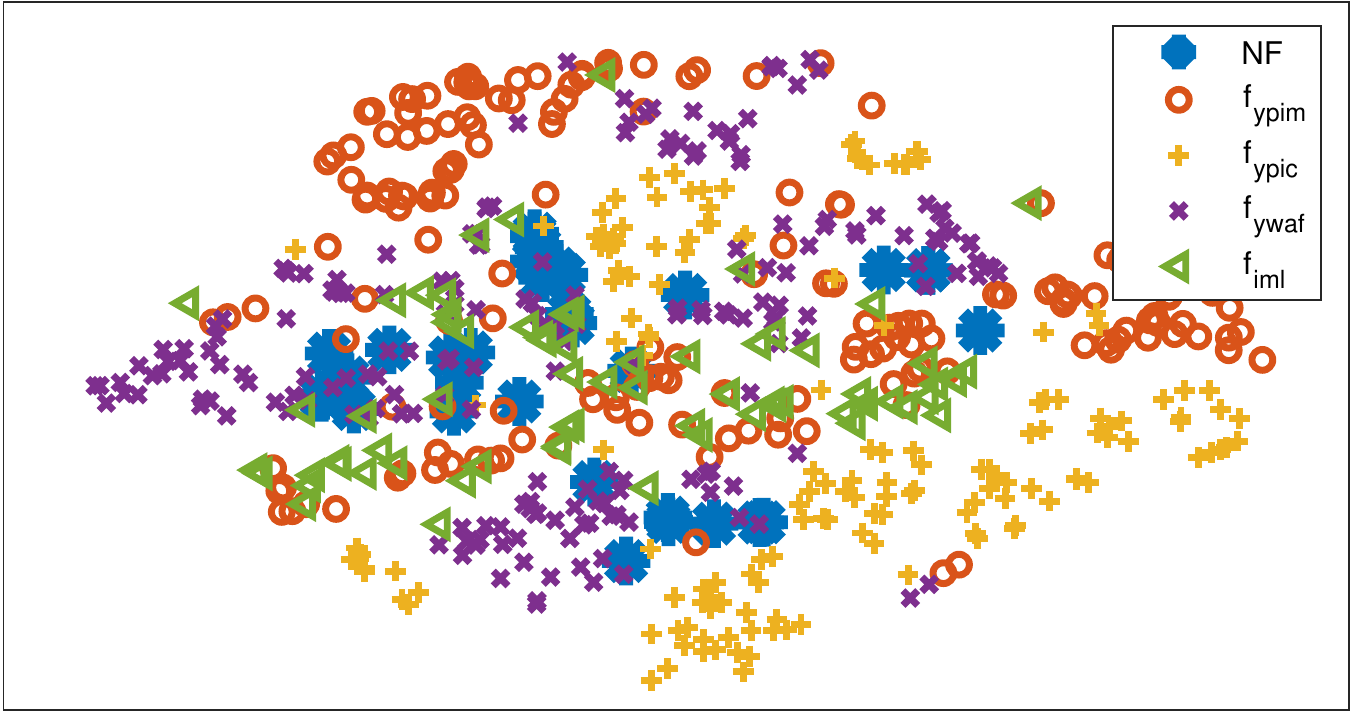}
    \caption{t-SNE plot of residual output data from the four residuals in \Figure\ref{fig:r_histograms} evaluated on different fault classes. Data is down-sampled to improve visualization.}
\end{figure}

\subsection{Prediction Error Estimation Using Ensemble Neural Networks}

Since training of neural networks is a non-linear optimization problem, 
it is beneficial to train multiple instances of each RNN and then select the model 
with best accuracy to reduce the risk of getting stuck in a bad local minima. Another 
way of using the multiple trained instances is to estimate confidence intervals. 

\begin{figure}
    \centering
    \includegraphics[width=1\columnwidth]{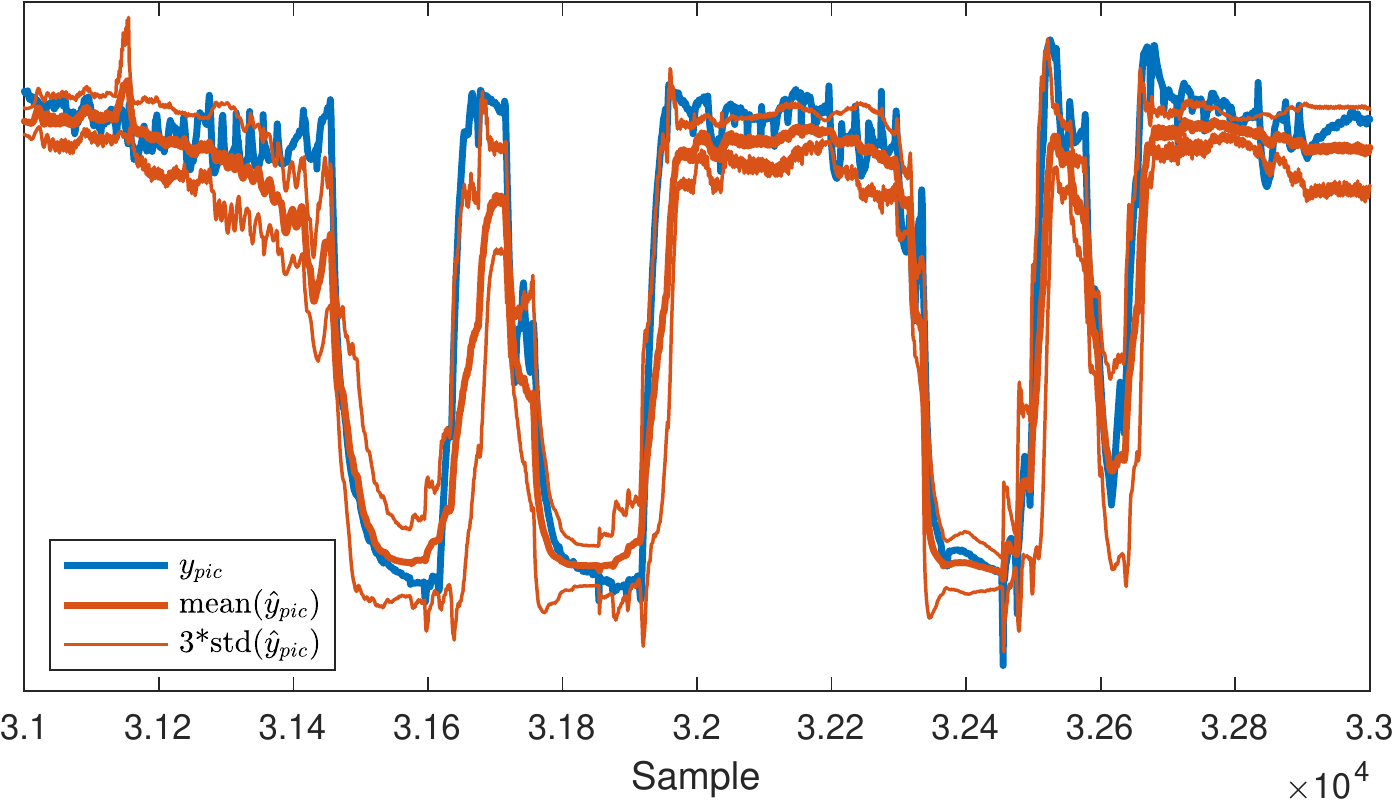}
    \caption{Ensemble prediction of sensor output $y_{pic}$ using eight trained neural 
    networks based on MSO$_{55}$. The plot show 
    mean prediction and 3*std confidence intervals. }
    \label{fig:mtes55_69_ensemble_prediction_nf}
\end{figure}

Results in \cite{lakshminarayanan2017simple} indicate that using random initialization of the 
parameters when training each neural network is sufficient to get a set of uncorrelated 
models. The eight trained instances are combined into an ensemble neural network 
prediction model. The ensemble model prediction is computed based on the 
average predicted value for all models and a confidence interval can be estimated by 
computing the variance over all predictions in each time instance. 
Figure~\ref{fig:mtes55_69_ensemble_prediction_nf} shows the mean and confidence 
intervals computed as 3*std based on the original predictions. The figure shows that the 
confidence intervals varies over time and, in general, increase during transients 
and are smaller during stationary operation. 

\subsection{Change Detection Using CUSUM Tests}

\begin{figure}
    \centering
    \includegraphics[width=1\columnwidth]{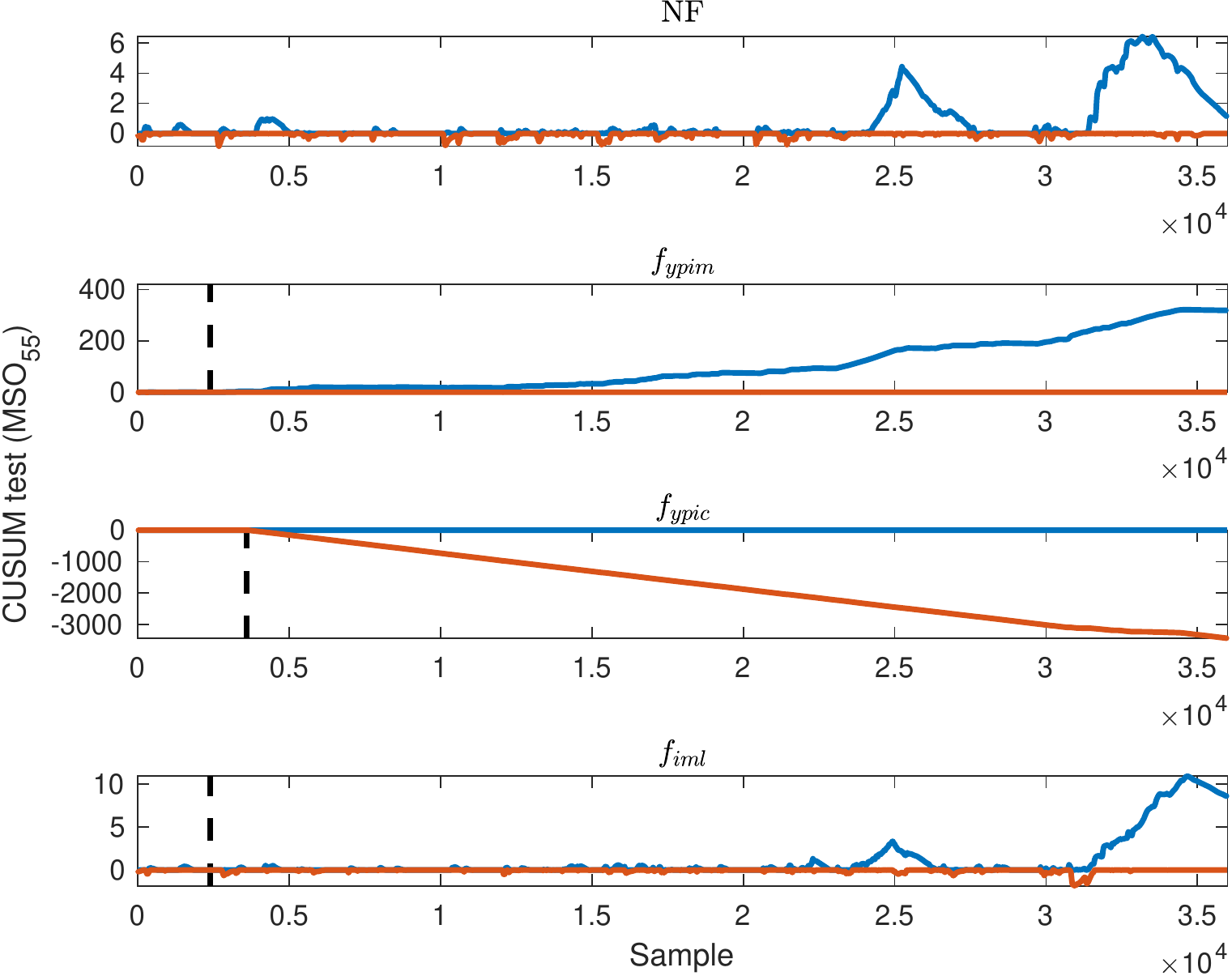}
    \caption{Fault detection using CUSUM tests based on residual generator MSO$_{55}$. The blue curve shows $T^+_t$ and the red curve $T^-_t$. 
    The dashed lines show when the faults occur in each scenario.}
    \label{fig:cusum_mtes55_69}
\end{figure}

\begin{figure}
    \centering
    \includegraphics[width=1\columnwidth]{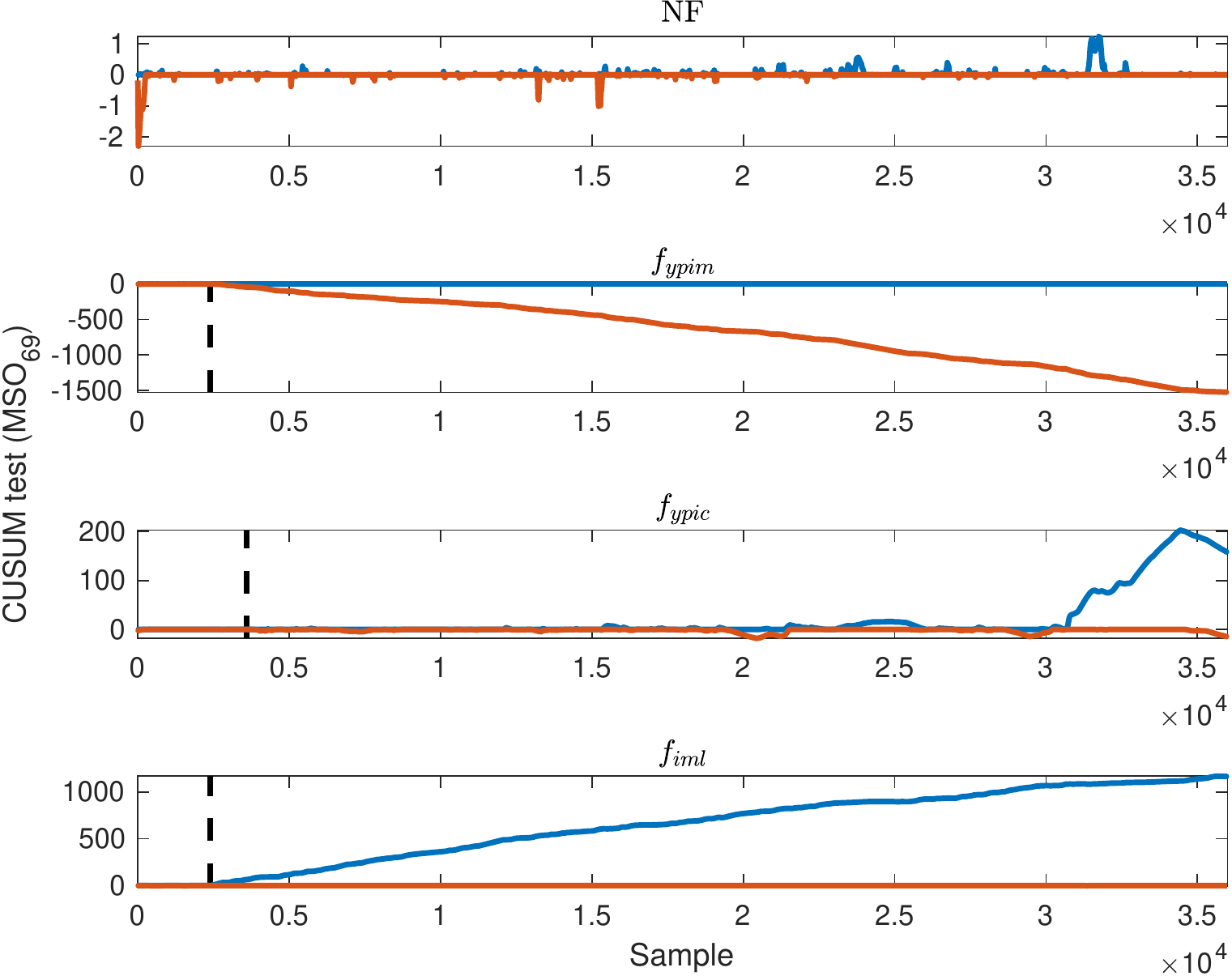}
    \caption{Fault detection using CUSUM tests based on residual generator MSO$_{69}$.}
    \label{fig:cusum_mtes69_71}
\end{figure}

The detection performance in \Figure\ref{fig:auc} is evaluated sample-by-sample. However, 
faults are usually present during a longer time period and, therefore, it is beneficial to take residual 
time-series information into consideration during the fault detection process. To improve 
classification performance, a set of CUSUM tests \eqref{eq:cusum} are constructed based 
on the residual outputs. CUSUM tests are used in, e.g., model-based diagnosis to improve 
fault detection performance of small faults by allowing a longer time to detection without 
increasing the risk false-alarms. For each residual, two CUSUM tests are created: one to 
detect positive change and one to detect negative change: 
\begin{equation}
\begin{aligned}
T^+_t &= \max\left(0, T^+_{t-1} + r_t - \nu^+ \right) \\
T^-_t &= \max\left(0, T^-_{t-1} - r_t - \nu^- \right)
\end{aligned}
\end{equation}
The parameters $\nu^+$ and $\nu^-$ in \eqref{eq:cusum} are tuned for each 
test using fault-free data to fulfill requirements regarding false alarm rate. 
The results in \Figure\ref{fig:cusum_mtes55_69} and \Figure\ref{fig:cusum_mtes69_71} 
show the CUSUM tests based on residual generators MSO$_{55}$ and MSO$_{69}$ respectively. 
The CUSUM tests are small in the nominal case and increases over time when 
the fault is present. Note that even though the AUC curve in \Figure\ref{fig:auc} 
is small, for example detection of fault $f_{ypim}$ using MSO$_{55}$, the CUSUM 
test accumulates fault information over time which improves detection performance
as shown in \Figure\ref{fig:cusum_mtes55_69}.   

\section{Discussion}

The engine case study show that structural methods is a useful approach to design 
grey-box RNN for residual generation. It is also shown how model-based approaches, such as
CUSUM tests, can improve classification performance. Residual generators are 
anomaly classifiers modeling nominal system operation represented in training data. Since 
an anomaly classifier only detects when observed data deviates from nominal training data, 
there are some important aspects to take into consideration when drawing conclusions from 
triggered residuals, i.e. residual outputs that deviate from the nominal case.

\subsection{Training Data}

The accuracy of a data-driven classifier depends on the quality of training data. 
If training data are not representative of nominal operation the neural network-based  
residual generators are not only likely to trigger an alarm when there is a fault in 
the system but also when the current system operation is not represented in training 
data. This means that a triggered residual can be explained by a fault or that the 
system is operating in a different mode with respect to training data.  

\subsection{Fault sensitivity}

Another known problem is sufficient excitation of all signals in training data. Evaluation of 
fault detection performance show that the trained grey-box RNN are overall best at 
detecting faults strongly affecting the predicted sensor output, even though they are 
sensitive to the other faults as well. Some 
measured states such as temperatures are slowly varying over time which requires 
more extensive data collection to cover the expected variations during nominal operation
Making sure that training data is representative of all relevant system operation is 
necessary to reduce the risk of false alarms but also assure that residuals become 
sensitive to faults affecting those signals. One solution is to design different residuals 
with varying sampling rates to capture fast and slow dynamics and, thus, be able to 
detect different types of faults.     

\subsection{Improve Fault Isolation Performance Using Data-Driven Fault Class Modeling}

Even though the residual generators are trained using only nominal training data, 
it is possible that data from different fault scenarios are available or will be collected 
over time. Thus, it is relevant to take this data into consideration to improve fault 
classification performance when new faults are detected. 

It is possible to distinguish between the two faults because they are affecting the residuals in 
different ways. Feeding the residual outputs into an multi-class classifier to compute fault 
hypotheses that can explain the observed data. Open set classification approaches
proposed in, e.g., \cite{jung2018combining} and \cite{scheirer2012toward}, could be 
used to handle unknown fault classes and incremental learning of fault models to improve 
classification performance over time.

\section{Conclusions}
Unknown fault classes and limited training data complicate data-driven 
fault diagnosis. In addition, developing sufficiently accurate models for 
designing model-based residuals is a difficult and time consuming process.  
Using grey-box RNN for residual generation is shown to be a 
promising approach to take advantage of both model-based 
fault diagnosis methods and data-driven model training. The use of grey-box 
RNN makes it possible to incorporate available physical insights about 
the system into data-driven models where tools from model-based diagnosis 
like structural analysis can help automate the residual generator design process. 
The results from the engine case study show that the generated grey-box RNN 
are able to capture the dynamic behavior of the engine and can be used to 
detect and classify different types of faults. 

Future work will further investigate how the model structure of the 
different grey-box neural networks affect their fault detection performance. 
It is also important to investigate how to optimize fault sensitivity when training 
the grey-box RNN to maximize fault detection but also make sure that faults 
are decoupled according to the structural model.

\section*{Acknowledgment}

The author would like to acknowledge Andreas Lundgren for his help with data collection. 
Computations have been performed at the Swedish National Super Computer Center (NSC). 
\ifCLASSOPTIONcaptionsoff
  \newpage
\fi



\bibliographystyle{IEEEtran}
\bibliography{references}
%

%





\end{document}